\documentclass[epj]{svjour}
\usepackage{graphics}
\begin{document}
\title{Searches for New Particles}
\author{Arnulf Quadt\inst{1}
}                     
\offprints{Arnulf.Quadt@cern.ch}          
\institute{Physikalisches Institut, Universit\"at Bonn, 
           Nu{\ss}allee 12, 53115 Bonn}
\date{Received: date / Revised version: date}
%
\abstract{Searches for New Particles including future sensitivity prospects 
are reviewed. The main focus is placed on results obtained at LEP,
HERA and the Tevatron on generic searches such as searches for excited
fermions, searches for leptoquarks and high-$p_T$ leptons. Also
interpretations in the context of anomalous single top production via
flavour-changing neutral current, large extra space dimensions,
supersymmetry, and various searches for Higgs bosons are discussed.
} 
\maketitle
\section{Introduction}
\label{sec:1}
In this review searches for new particles are summarized. The main
focus is placed on results from current data as well as future
expected sensitivities, where relevant, from LEP, HERA and the
Tevatron. This review covers direct searches for particles in and
beyond the Standard Model, while indirect constraints on physics
beyond the Standard Model from precision electroweak data are mostly
described in \cite{eps_pippa}. Future prospects on searches for new
particles at the LHC and/or a linear collider (LC) are summarized in
more detail in \cite{lhc_lc_summaries}. 

Selected topics and recent results are being reviewed. Firstly two
examples of generic searches, motivated by the observed structure in
the fermion sector are discussed, namely the search for excited
fermions and the search for leptoquarks, followed by searches for
anomalous single-top production, large extra space dimensions,
searches for supersymmetry and various searches for Higgs bosons
\footnote{Unless explicitly specified all exclusions limits are quoted
at 95\% CL.}.

\section{Excited Fermions}
\label{sec:2}
Charged $(e^*, \mu^*, \tau^*)$ and neutral $(\nu_e^*, \nu_\mu^*,
\nu_\tau^*)$ excited leptons are predicted by composite models where
leptons and quarks have substructure \cite{exferm_models}. These
models address fundamental questions left open by the Standard Model,
such as the number of three observed fermion families, the hierarchy
in the fermion mass values, and the similarity in the electric charge
and weak properties of the fermions.  A consequence of the possible
fermion compositeness or substructure would be the existence of
excited fermion states with masses of the exited state in the order
of $100\;\rm GeV$, which are expected to be accessible to various
experiments. The most commonly use phenomenological model
\cite{exferm_models} is based on the assumption that the excited
fermions have spin and isospin $\frac{1}{2}$ and both left-handed,
$F_L^*$, and right-handed components, $F_R^*$, are in a weak
isodoublets. The Lagrangian describes the transitions between known
fermions, $F_L$, and excited states:
\begin{eqnarray}
\cal{L}_{F^*F} &=&
\frac{1}{\Lambda} \overline{F_R^*} \sigma^{\mu \nu} \left[ g\, f\,
\frac{\vec{\tau}}{2} \partial_\mu \vec{W_\nu} + g^\prime\, f^\prime\,
\frac{Y}{2} \partial_\mu B_\nu  + \right.\\
  & & \phantom{\frac{1}{\Lambda} \overline{F_R^*} \sigma^{\mu \nu} 
      \left[ \right.} g_s\, f_s\,  \left.
\frac{\lambda^a}{2} \partial_\mu G_\nu^a \right] F_L + h.c. \nonumber
\end{eqnarray}
where $\Lambda$ is the compositeness scale; $\vec{W_\nu}, B_\nu$ and
$G_\nu^a$ are the SU(2), U(1) and SU(3) fields; $\vec{\tau}, Y$ and
$\lambda^a$ are the corresponding gauge-group generators; and $g,
g^\prime$ and $g_s$ are the coupling constants. The free para\-me\-ters
$f, f^\prime$ and $f_s$ are weight factors associated with the three
gauge groups and depend on the specific dynamics describing the
compositeness. For an excited fermion to be observable, $\Lambda$ must
be finite and at least one of $f, f^\prime$ and $f_s$ must be
non-zero. By assuming relations between $f, f^\prime$ and $f_s$, the
branching ratios of the excited-fermion decays can be fixed, and the
cross section depends only on $f/\Lambda$ and on the mass of the
excited fermion, $m_{f^*}$.

\begin{figure}[ht]
\centerline{
\resizebox{0.36\textwidth}{!}{%
  \includegraphics{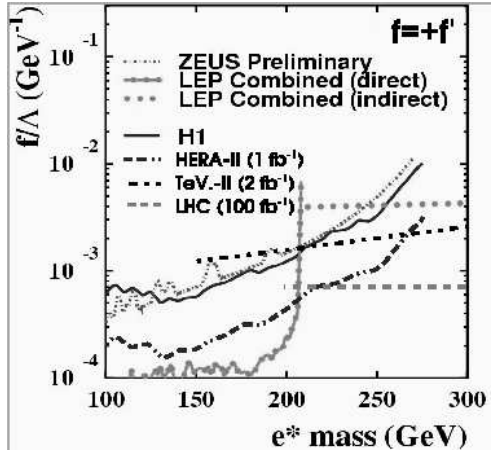} }}

\caption{Upper limits on the coupling 
         $f/\Lambda$ as a function of the excited lepton mass for HERA
         and LEP experiments together with the expected sensitivity
         for HERA-II, Tevatron-II and the LHC.}
\label{fig:1}       
\end{figure}

For the example of excited electrons, the conventional relation $f =
f^\prime$ is adopted. Searches for excited electrons have been carried
out at HERA in single $e^*$ production with subsequent decays to $e,
\nu$ and $\gamma, Z, W$ and at LEP in pair-production, single production,
and indirect via $t$-channel $e^*$-exchange, i.e. multi-photon events
\cite{estar_lep_hera}. At low $e^*$-masses up to the kinematic 
pair-production limit at LEP close to $100\;\rm GeV$ excited electrons
can be ruled out for all values of $f/\Lambda$ (Figure~\ref{fig:1}),
while in single $e^*$ production searches masses up to $\sqrt{s}$ can
be excluded for the coupling $f/\Lambda$ down to about $10^{-4}\;\rm
GeV^{-1}$.  Above $300\;\rm GeV$ the best direct exclusion limits have
been achieved by the HERA experiments while at masses beyond $250\;\rm
GeV$ the search for $t$-channel $e^*$ exchange in multi-photon events
provides strong indirect limits on $f/\Lambda$ to be below $\sim 5
\times 10^{-3}\;\rm GeV^{-1}$. Preliminary studies show that HERA-II
with $1\;\rm fb^{-1}$ will improve the HERA sensitivity by a factor
3-4, the Tevatron with $2\;\rm fb^{-1}$ is expected to have
sensitivity down to $2 \times 10^{-3}\;\rm GeV^{-1}$ and the LHC with
$100\;\rm fb^{-1}$ even down to $7\times 10^{-4}\;\rm GeV^{-1}$ in
$f/\Lambda$, also at high $e^*$ masses \cite{estar_tev_lhc}. Similar
to their $e^*$ results, the LEP and HERA experiments have also
obtained search results for $\mu^*, \tau^*$ and for $\nu^*$. In the
context of contact term production CDF has performed a search for
excited electrons with $72\;\rm pb^{-1}$ of Run-II data and excludes
masses $M_{e^*}$ up to $785\;\rm GeV$ for $M_{e^*} / \Lambda = 1$.

\begin{figure}[ht]
\centerline{
\resizebox{0.30\textwidth}{!}{%
  \includegraphics{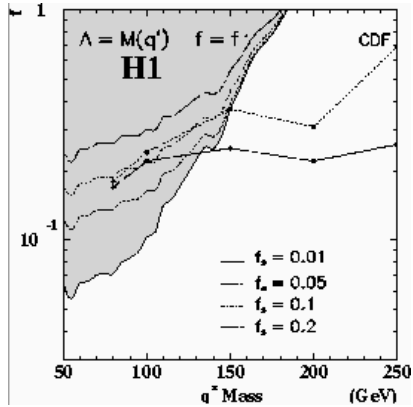}
}}
\caption{Upper limits on the coupling 
         $f$ as a function of the excited quark mass for the H1 and
         CDF experiments for different choices of the relative strong
         coupling $f_s$.}
\label{fig:2}       
\end{figure}

The search for excited quark production with de-ex\-ci\-ta\-tion via
radiation of $W, Z, \gamma,$ gluons is most sensitive at the hadron
colliders HERA and Tevatron \cite{qstar_search_exp}. For $f =
f^\prime$ HERA reaches, depending on the choice of $f_s$, a
sensitivity of down to $f = 0.06$ at low masses of $q^*$, while
studies of the dijet-mass spectra at the Tevatron provide higher
sensitivity at large $q^*$ masses (Figure~\ref{fig:2}). For $f =
f^\prime = f_s = 1$ and $\Lambda = M_{q^*}$ D{\O} in Run-I and CDF in
Run-II reach exclusions of up to $775\;\rm GeV$ and $760\;\rm GeV$,
respectively, while a mass exclusion sensitivity for $2\;\rm fb^{-1}$
data of up to $940\;\rm GeV$ is expected. In the coming years further
sensitivity improvements are expected from the continuing data taking
of the hadron collider experiments. Further details of searches for
excited fermions can be found in \cite{eps2003_excited_fermions}.

\section{Leptoquarks}
\label{sec:3}
The known symmetry between the lepton and quark sectors could possibly
be an indication that they are fundamentally connected through a new
interaction. Models for such interactions predict the existence of
Leptoquarks (LQs) which are colour-triplet scalar (S) or vector (V)
bosons carrying lepton and baryon numbers, and a fractional
electromagnetic charge, $Q_{em}$. Most collider searches are carried
out in the context of effective models, in particular in the
Buchm\"uller, R\"uckl and Wyler (BRW) model \cite{brw_model}. This
model describes leptoquark interactions in a most general effective
Lagrangian under the assumption that leptoquarks have renormalizable
interactions invariant under SM gauge groups, and that they couple
only to gauge bosons and to ordinary fermions. Further assumptions are
that leptoquarks conserve leptonic number and baryonic number
separately (flavour-diagonal). These assumptions result in leptoquark
decay branching ratios to electrons and quarks of 0, 50\% or 100\%.

First-generation leptoquarks can be resonantly produced at HERA by the
fusion of an $e$ beam particle with a $q$ from the proton. This
process interfers with $t$-channel electroweak-boson exchange. At the
Tevatron leptoquarks can be pair-produced, independent of the
leptoquark coupling $\lambda$ to electrons and quarks, with subsequent
decays to charged lepton or neutrino and a quark. At LEP leptoquarks
can be virtually exchanged in the $t$-channel and result in di-jet
final states.

\begin{figure}[ht]
\centerline{
\resizebox{0.35\textwidth}{!}{%
  \includegraphics{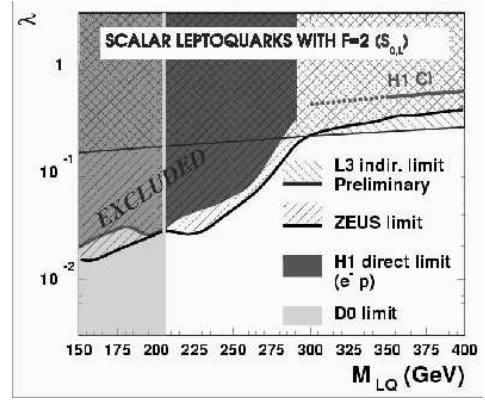}
}}
\caption{Constraints from HERA experiments, LEP and D{\O} 
         in the framework of the BRW model for the example of a
         leptoquark with fermion number $F=2$.}
\label{fig:3}       
\end{figure}

Figure~\ref{fig:3} shows the results from HERA, LEP and the Tevatron
on leptoquark searches \cite{eps03_leptoquark}, in particular choosing
here a typical scalar with $F=2$, namely the $S_{0,L}$ for which
$\beta_{eq} = Br(LQ \rightarrow eq) = 0.5$. The Tevatron (Run-I)
exclusion limit of $204\;\rm GeV$ is independent of $\lambda$. For
$\lambda \ll 1$, in the mass range beyond the Tevatron reach and below
$\sim 300\;\rm GeV$, HERA has the highest sensitivity down to
$\lambda$ of a few $\times 10^{-2}$ which will be superseeded by
HERA-II. For leptoquark masses larger than the HERA $\sqrt{s}$ the
virtual leptoquark exchange at HERA and LEP provides similar
sensitivity to about the electromagnetic strength ($\lambda = \sqrt{4
\pi \alpha} \approx 0.3$).

\begin{figure}[ht]
\centerline{
\resizebox{0.49\textwidth}{!}{%
  \includegraphics{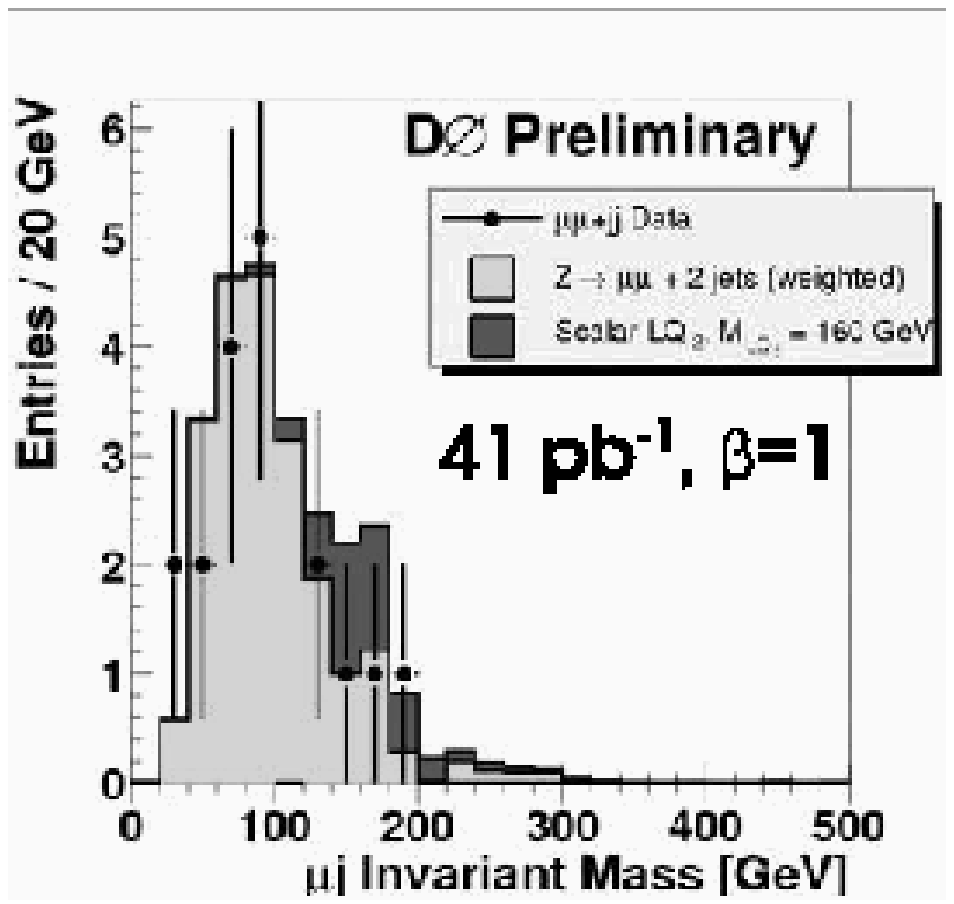}
  \includegraphics{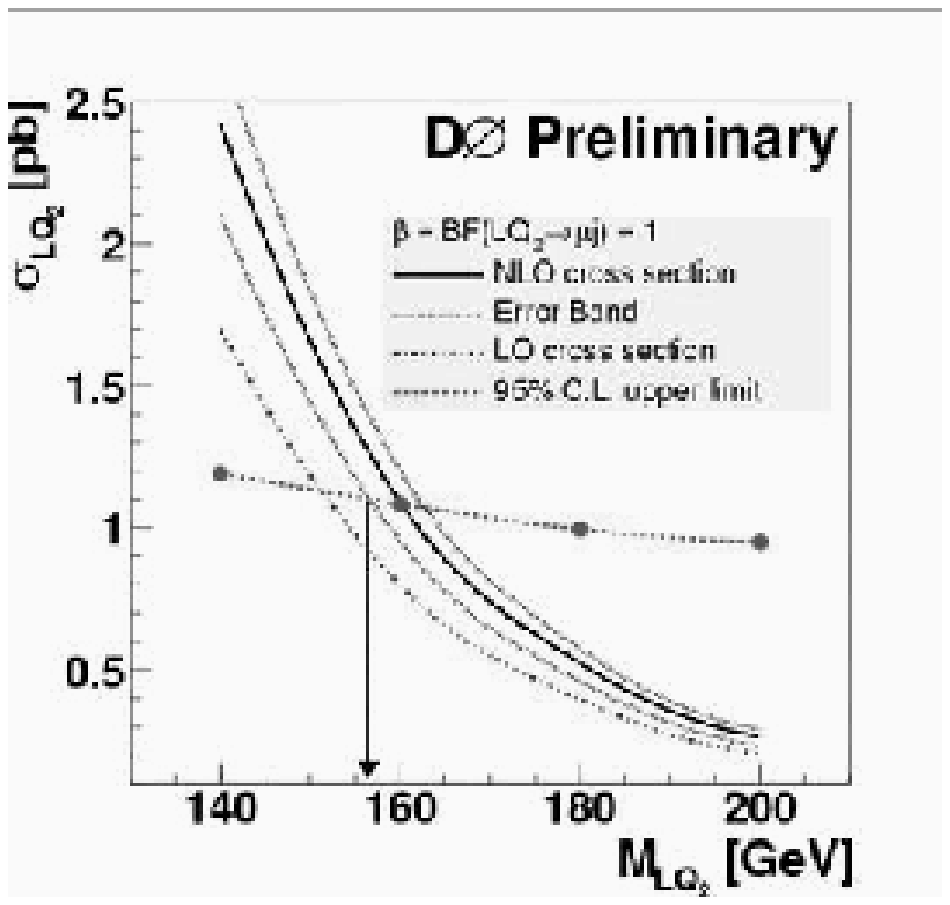}
}}
\caption{Left: Invariant $\mu q$-mass distribution for early D{\O} 
         Run-II data in comparison to dominant background and LQ
         signal. Right: Expected LQ cross section and excluded cross
         section as a function of LQ mass.}
\label{fig:4}       
\end{figure}

The Tevatron experiments CDF and D{\O} have already performed searches
for scalar leptoquarks in the early Run-II data
\cite{eps03_leptoquark}. In addition to the already established
searches for 1st generation leptoquarks, D{\O} has also searched for
2nd generation leptoquarks. Figure~\ref{fig:4} shows the resulting
invariant $\mu q$ mass distribution with a hypothetical signal of
$160\;\rm GeV$ along with the expected signal cross section and the
experimentally excluded cross section as a function of leptoquark
masses. While the typical mass exclusions in the different channels
range from $107\;\rm GeV$ to $230\;\rm GeV$ (Table~\ref{tab:1}), they
are expected to increase with $2\;\rm fb^{-1}$ of data to $250 -
325\;\rm GeV$ in the future. Further details of the leptoquark
searches can be found in
\cite{eps03_leptoquark}.

\begin{table}[ht]
\caption{Preliminary Tevatron Run-II mass exclusions
         in the different channels for 1st and 2nd generation leptoquarks.}
\label{tab:1}       
\centerline{
\begin{tabular}{lrl}
\hline\noalign{\smallskip}
experiment & channel (lumi) & $M_{LQ} >$ \\
\noalign{\smallskip}\hline\noalign{\smallskip}
CDF  & $eeqq       \, (72\,\rm pb^{-1})$ & 230 GeV\\
     & $\nu\nu qq  \, (76\,\rm pb^{-1})$ & 107 GeV\\
D{\O}& $eeqq       \, (41\,\rm pb^{-1})$ & 179 GeV\\
     & $\mu \mu qq \, (41\,\rm pb^{-1})$ & 157 GeV\\
\noalign{\smallskip}\hline
\end{tabular}}
\end{table}

\section{High \boldmath{$p_t$}-Leptons with Large Missing Transverse Momentum and 
         Limits on Anomalous Top Couplings}
\label{sec:4}
\begin{figure}[ht]
\centerline{
\resizebox{0.24\textwidth}{!}{%
  \includegraphics{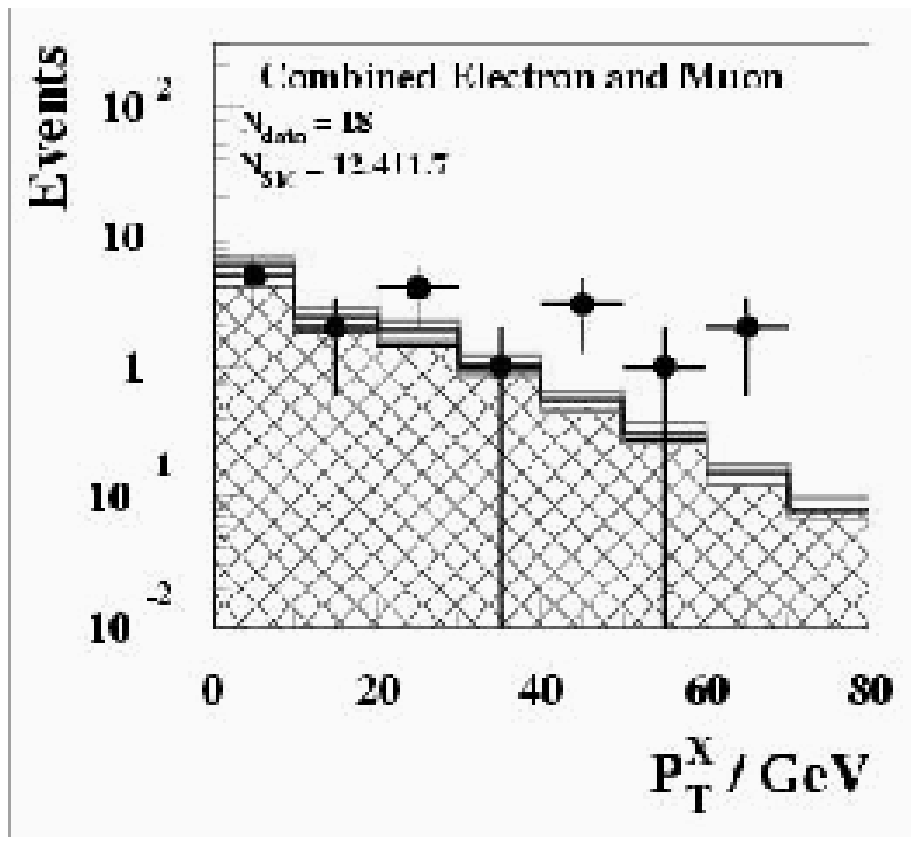}
}
\resizebox{0.23\textwidth}{!}{%
  \includegraphics{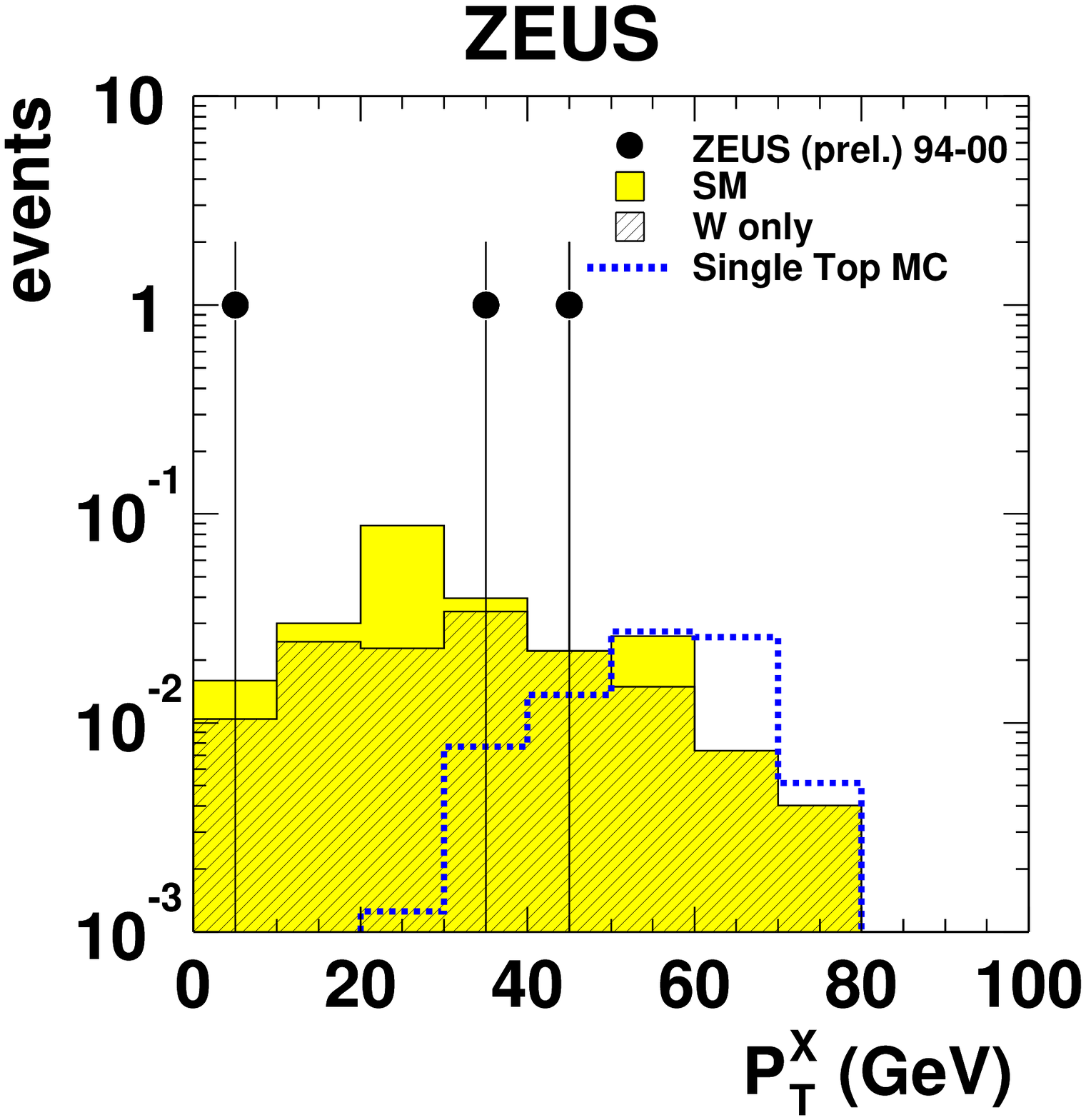}
}}
\caption{Distributions of the hadronic transverese momentum 
         $p_T^X$ combined in the electron and muon channel for H1
         (left) and in the tau channel for ZEUS (right).}
\label{fig:5}       
\end{figure}

At the HERA experiments H1 and ZEUS searches for events containing
isolated high-$p_t$ leptons and large missing transverse momentum have
been performed \cite{hera_highpi_met}. A previously reported excess of
events in the electron and muon channel by H1 and in the tau channel
by ZEUS (Table~\ref{tab:2}, Figure~\ref{fig:5}) has been discussed in
the context of potential anomalous flavour-changing neutral current
(FCNC) single-top production.

\begin{table}[ht]
\caption{Observed and expected number of high-$p_t$ lepton 
         events with large missing transverse momentum for H1
         ($118.3\;\rm pb^{-1}$) and for ZEUS ($130.1\;\rm pb^{-1}$).}
\label{tab:2}       
\centerline{
\begin{tabular}{rrrr}
\hline\noalign{\smallskip}
H1  & $e$ obs./exp. & $\mu$ obs./exp. & combined \\
\noalign{\smallskip}\hline\noalign{\smallskip}
$p_T^X > 25\;\rm GeV$ & $4 / 1.49 \pm 0.28$ & $6 / 1.44 \pm 0.26$ & $10 / 2.93 \pm 0.49$ \\
$p_T^X > 40\;\rm GeV$ & $3 / 0.54 \pm 0.11$ & $3 / 0.55 \pm 0.12$ & $ 6 / 1.08 \pm 0.22$ \\
\noalign{\smallskip}\hline\noalign{\smallskip}
ZEUS  &  &  & $\tau$ obs./exp. \\
\noalign{\smallskip}\hline\noalign{\smallskip}
$p_T^X > 25\;\rm GeV$ & $2 / 2.90 \pm 0.59$ & $5 / 2.75 \pm 0.21$ & $ 2 / 0.12 \pm 0.02$ \\
$p_T^X > 40\;\rm GeV$ & $0 / 0.94 \pm 0.11$ & $0 / 0.95 \pm 0.14$ & $ 1 / 0.06 \pm 0.01$ \\
\noalign{\smallskip}\hline
\end{tabular}}
\end{table}

In the SM the neutral currents are flavour diagonal. Flavour-changing
neutral currents (FCNC) are not contained at tree level and can happen
only from higher-order loop contributions. Sizeable rates can arise
only when the top quark appears in the loop. Therefore, no detectable
rate is predicted in the SM for FCNC processes between the top and
charm or up quarks. However, considerable enhancements are expected,
especially at large FCNC \cite{kuze_sirois}, in various new models
such as models with two or more Higgs doublets, supersymmetric models
with or without $R$-parity conservation, or models with a composite
top quark. The top quark phenomenology is less tightly constrained
than that of lighter quarks and can be tested at current colliders. In
the absence of a specific predictive theory, a most general effective
Lagrangian was proposed to describe FCNC top interactions involving
electroweak bosons.

At HERA single-top production can proceed via $\gamma$ or $Z$
$t$-channel exchange through anomalous FCNC couplings
$\kappa_{tq\gamma}$ or $v_{tqZ}$, where $q$ is a $u$- or a
$c$-quark. The sensitivity to couplings for the $u$-quarks is much
higher than that for the $c$-quark because of the smallness of the
$c$-quark density in the proton, the sensitivity to $\kappa$ is higher
than that to $v$ due to the propagator suppression in the $t$-channel
$Z$ exchange. In comparison to previous analyses the simulation of the
$Z$-exchange process has now been improved allowing an increased
sensitivity to FCNC couplings at large $v_{tuZ}$.

The single-top production at HERA would yield a high-$p_T$ $W$-boson
accompanied by an energetic $b$-quark jet. When the $W$ decays
leptonically the event topology will contain an energetic isolated
lepton and large missing transverse momentum, as well as large
hadronic transverse momentum, i.e. the topology observed in excess of
the SM expectation. For the hadronic decay of the $W$, the topology
will be a three-jet event with a resonant structure in dijet and
three-jet invariant masses.

At LEP FCNC single-top quark production has been searched for in the
$e^+ e^- \rightarrow tc(u) \rightarrow bWc(u)$ process, where hadronic
and leptonic decay modes of the $W$ are considered
\cite{kuze_sirois,lep_t}. The LEP experiments have roughly similar
sensitivity to $\kappa_\gamma$ and $v_Z$, which increases with lower
top quark mass.  At the Tevatron, CDF has performed a search for FCNC
in the top decays $t \rightarrow \gamma c(u)$ and $t \rightarrow Z
c(u)$ in $p\bar{p}$ collisions \cite{kuze_sirois,cdf_t}. The resulting
limits on the top quark decay branching ratios have been converted
into limits on the FCNC couplings $\kappa_\gamma$ and $v_Z$.

\begin{figure}[ht]
\centerline{
\resizebox{0.30\textwidth}{!}{%
  \includegraphics{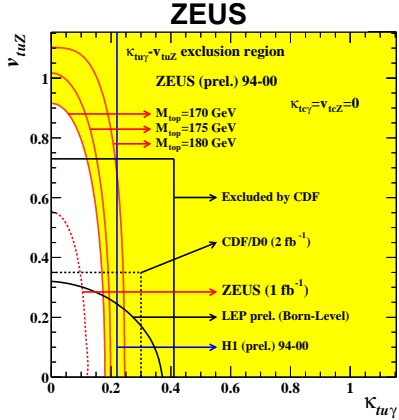}}
}
\caption{Exclusion regions in the $\kappa_{tu\gamma} - v_{tuZ}$ 
         plane as achieved by the different experiments along with the
         expected sensitivity for future data taking at HERA and the
         Tevatron.}
\label{fig:6}       
\end{figure}

Figure~\ref{fig:6} shows the exclusion regions in the
$\kappa_{tu\gamma} - v_{tuZ}$ plane as achieved by the different
experiments along with the expected sensitivity for future data taking
at HERA and the Tevatron\footnote{Note that the Lagrangian used by LEP
and by HERA experiments differ by a multiplicative factor such that
$\kappa^{LEP}_{tu\gamma} = \sqrt{2} \, \kappa^{ZEUS}_{tu\gamma}$ and
$v^{LEP}_{tuZ} = \sqrt{2} \, v^{ZEUS}_{tuZ}$; also note that LEP and
Tevatron are sensitive to $u$- as well as $c$-quark couplings with
equal strength.}. LEP has the highest sensitivity to $v_{tuZ}$ which
is expected to become similar for the Tevatron after Run-IIA
\cite{fcnc_tevii}, whereas HERA-I has the highest sensitivity to
$\kappa_{tu\gamma}$, which will be improved even further with HERA-II
\cite{fcnc_heraii}. The observed excess of events in particular by H1
results in slightly weaker exclusion limits in $\kappa_{tu\gamma}$
compared to ZEUS. The tau events by ZEUS are unlikely to be explained
by the hypothesis of single top quark production and are not included
in this exclusion plot on FCNC couplings.

\section{Large Extra Space Dimensions}
\label{sec:5}

The Standard Model of particle physics is an extraordinary scientific
achievement, with nearly every prediction confirmed to a high degree
of precision. Nevertheless, the SM still has unresolved and
unappealing characteristics, including the problem of a large
hierarchy in the gauge forces, with gravity being a factor of $10^{33}
- 10^{38}$ weaker than the other three forces, raising the question
why $M_{Pl} / m_{EW} \sim 10^{15}$ is so large. Viable quantum-gravity
scenarios have been constructed \cite{add,rs} in which the
gravitational force is expected to become comparable to the gauge
forces close to the EW scale, eventually leading to (model dependent)
effects in the TeV range, observable at high energy colliders
\cite{grw}: virtual graviton exchange and direct graviton emission.

In the string-inspired Arkani-Hamed, Dimopoulos and Dvali (ADD)
scenario \cite{add} whith $n \ge 2$ (6 in string theories) ``large''
compactified extra dimensions, a gravitational ``string'' scale $M_s$
is introduced in $(4 + n)$ which is related to the usual (effective
four-dimensional) Planck scale via $M_{Pl}^2 = R^n\, M_s^{2+n}$, where
$R$ is a characteristic (large) size of the $n$ compactified extra
dimensions. The gravitons are allowed to propagate in these extra
dimensions of finite size $R$ which implies that it will appear in our
familiar 4-dimensional universe as a ``tower'' of massive Kaluza-Klein
(KK) excitation states. The exchange of KK towers between SM particles
leads to an effective contact interactions with a coupling coefficient
$\eta_G = \lambda/M_s^4$. The SM fields are localized to the
4-dimensional space-time. For $r \ll R$ the gravitational potential
results from Gauss's law in $(n+4)$ dimensions while for $r \gg R$ it
takes the conventional form $V \sim 1/r$. Typical compactification
radii are:
\begin{table}[ht]
\label{tab:nn}       
\centerline{
\begin{tabular}{lll}
$n = 1$ & $R \sim 10^{13}\;\rm cm$ & empirically excluded\\
$n = 2$ & $R \sim 100\;\rm \mu m - 1\;\rm mm$ & under investigation\\
$n = 3$ & $R \sim 3\;\rm nm$ 
\end{tabular}}
\end{table}

In the Randall-Sundrum \cite{rs} (RS) model only one compact extra
dimension is introduced. In a similar way to the ADD model, SM
particles and forces are constrained to the 4-dimensional SM
brane. Only gravity is allowed to propagate into the extra
dimension. In this model the hierarchy is not generated by the extra
volume, but by a specifically chosen geometry (`warped' geometry). As
a direct consequence of the geometry, gravity is mainly located in a
distance $r_0$ at a second brane, the Planck brane, and propagates
exponentially damped into the extra dimension. Thus, there is only a
small overlap between gravity and SM particles and forces, explaining
the weakness of gravity with respect to the electroweak interaction,
or the observed mass hierarchy.

\begin{figure}[ht]
\centerline{
\resizebox{0.35\textwidth}{!}{%
  \includegraphics{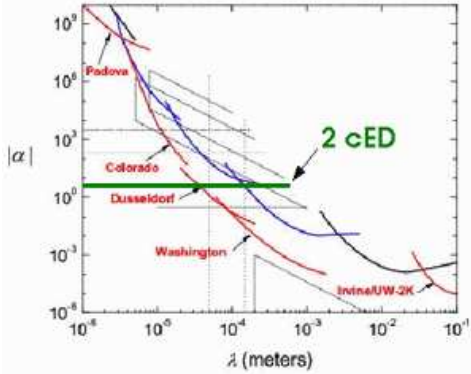}}
}
\caption{Current limits on new gravitational strength forces between 
         $1\;\rm \mu m$ and $10\;\rm cm$. Shown are the achieved (blue)
         as well as the in the future expected (red) limits on the
         Yukawa strength $\alpha$ as a function of the range
         $\lambda$.}
\label{fig:7}       
\end{figure}

Experiments in which deviations from Newton's gra\-vi\-ta\-tion law
are investigated at short distances usually consider the combined
potential energy V due to a modulus force and newton gravity to be
written as:
\begin{eqnarray}
V = - \int d\vec{r_1} \int d\vec{r_2}\, & & 
         \frac{G \rho_1(\vec{r_1}) \rho_2(\vec{r_2})}{r_{12}} \times \\
& & \left[ 1 + \alpha \, \exp (-r_{12} / \lambda)  \right] \nonumber
\end{eqnarray}
with $G$ the gravitational constant, $r_{12}$ the distance between two
points $\vec{r_1}$ and $\vec{r_2}$ in the test masses, and $\rho_1,
\rho_2$ the mass densities of the two bodies. $\alpha$ is the strength
of the new Yukawa force relative to gravity, and $\lambda$ the range.
The present limits on the Yukawa strength $\alpha$ as a function of
the range $\lambda$, as obtained from sophisticated short distance
gravity experiments \cite{shortrange_gravity}, is shown in
Figure~\ref{fig:7}. For $n=2$ large extra space dimensions new forces
with ranges below $\sim 200\;\rm \mu m$ are excluded.

\subsection{Indirect Effects - Virtual Graviton Exchange}
\label{sec:5.1}

\begin{figure}[ht]
\centerline{
\resizebox{0.41\textwidth}{!}{%
  \includegraphics{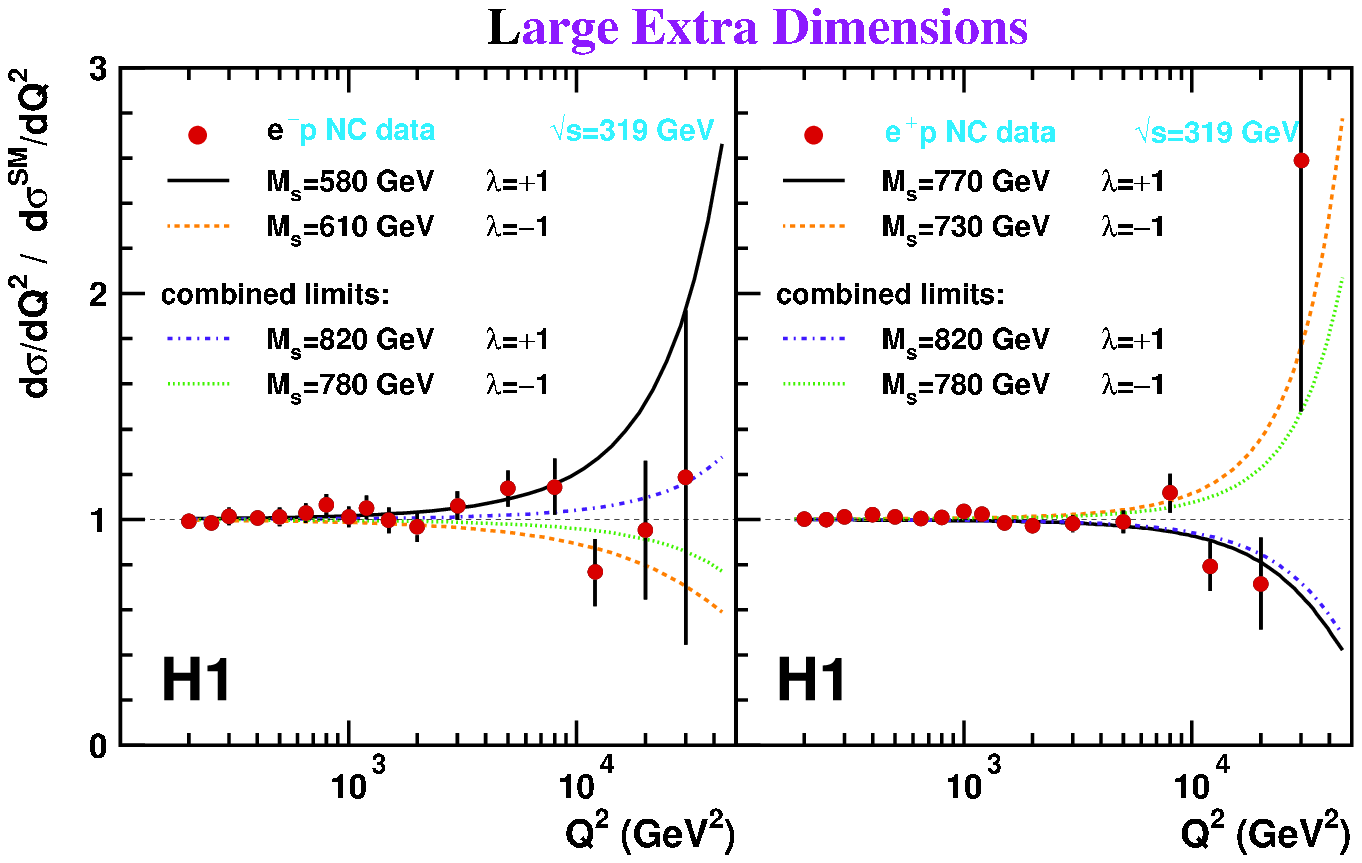}}
}
\caption{Constraints for models with large extra dimensions from 
         $Q^2$ distribution of NC-DIS events at HERA, here H1.}
\label{fig:8}       
\end{figure}

Searches for virtual graviton exchange in theories with large extra
dimensions have been performed at HERA, LEP and the Tevatron. At HERA
virtual graviton exchange modifies the $Q^2$ distribution of neutral
current events in a characteristic way. Figure~\ref{fig:8} shows the
ratio of high-$Q^2$ NC DIS events from H1 to the SM, together with the
effect of Kaluza-Klein graviton exchange for the just excluded
fundamental scale $M_s$. The coupling $\lambda$, which depends on the
full theory and is expected to be of order unity, has been fixed by
convention to $\lambda = \pm 1$. Combining both $\lambda = -1$ and
$\lambda = +1$ typical limits from H1 and ZEUS are $M_s > 0.8\;\rm
TeV$.

Indirect effects from virtual graviton exchange in large extra
dimensions at LEP are searched for in boson ($\gamma, Z$) and fermion
pair production, taking the modified mass and angular distribution of
the final state particles into account. Typical limits obtained for
the combination of the LEP data are $M_s > 1.26\;\rm TeV$ for $\lambda
= +1$ and $M_s > 0.96\;\rm TeV$ for $\lambda = -1$.

\begin{figure}[ht]
\centerline{
\resizebox{0.33\textwidth}{!}{%
  \includegraphics{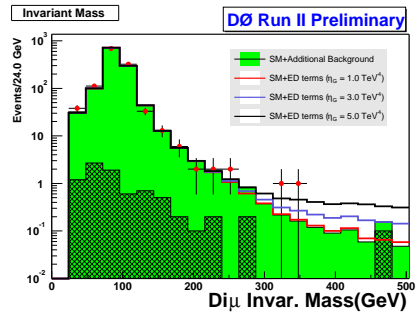}}
}
\caption{Comparison of invariant $\mu \mu$ mass distribution between 
         data ($30\;\rm pb^{-1}$ D{\O} data; points) and background
         (histograms) together with the modifications expected in the
         presence of large extra space dimensions at high $m_{\mu
         \mu}$.}
\label{fig:9}       
\end{figure}

At the Tevatron virtual graviton exchange, expected to modify the
invariant mass and angular distributions of the final state particles,
is searched for in boson ($\gamma, Z$) and fermion ($e, \mu$) pair
production (Figure~\ref{fig:9}). Typical limits obtained by CDF or
D{\O} are $M_s > 0.79 - 1.28\;\rm TeV$, depending of the channel and
data set (Run-I or Run-II).

\subsection{Direct Effects - Real Graviton Emission}
\label{sec:5.2}

Searches for direct effects of large extra dimensions, namely the real
graviton emission, have been performed at LEP and the Tevatron.  Real
gravitons can be emitted in association with photons or gluons,
resulting in large missing transverse momentum and a single photon or
gluon, i.e. a monojet.

\begin{figure}[ht]
\centerline{
\resizebox{0.45\textwidth}{!}{%
  \includegraphics{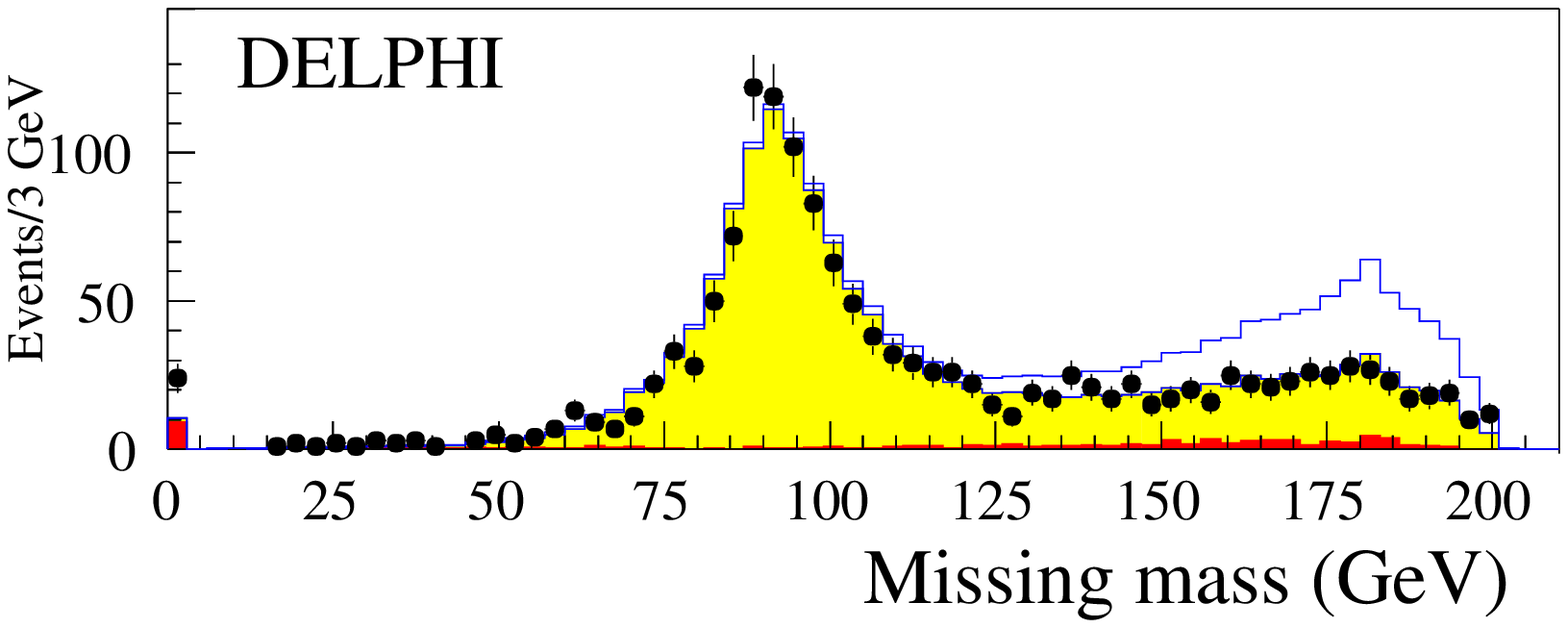}}
}
\caption{Missing mass (or recoil mass) distribution of all single photon 
         events in DELPHI. The light shaded area is the expected
         distribution from $e^+e^- \rightarrow \nu \bar{\nu}\gamma$,
         the dark shaded area is the total background from other
         sources. The expected signal $e^+e^- \rightarrow \gamma G$
         production is indicated.}
\label{fig:10}       
\end{figure}

\begin{figure}[ht]
\centerline{
\resizebox{0.35\textwidth}{!}{%
  \includegraphics{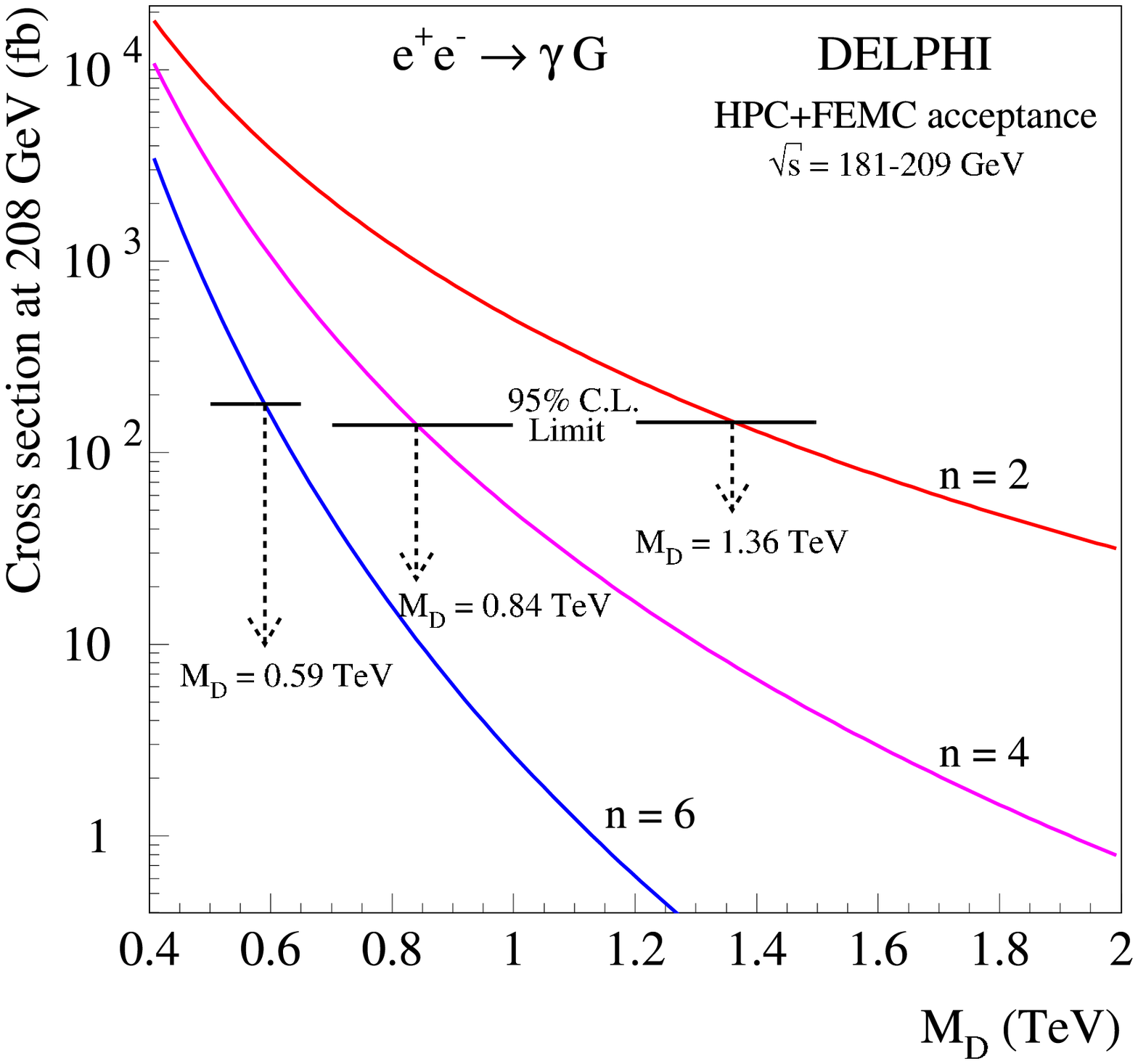}}
}
\caption{DELPHI limit and expected cross section for 
         $e^+e^- \rightarrow \gamma G$ for 2, 4, and 6 extra
         dimensions.}
\label{fig:11}       
\end{figure}

Figure~\ref{fig:10} shows the missing mass (or recoil mass)
distribution of all single photon events in DELPHI (from data recorded
at $\sqrt{s} = 181 - 209\;\rm GeV$). The light shaded area is the
expected distribution from $e^+e^- \rightarrow \nu \bar{\nu}\gamma$
and the dark shaded area is the total background from other
sources. The expected signal $e^+e^- \rightarrow \gamma G (n=2, M_D =
0.75\;\rm TeV)$ production is indicated. This high statistics
measurement and the well controlled SM background allow to set cross
section limits on the $\gamma G$ production as a function of the
fundamental scale $M_D$. Figure~\ref{fig:11} shows the result for the
DELPHI experiment for $n = 2, 4$ and $6$ large extra dimensions along
with the expected cross section. For $n = 2$ extra dimensions $M_D >
1.36\;\rm TeV$ which corresponds to $R < 260\;\rm \mu m$ and for $n =
4$ extra dimensions $M_D > 0.84\;\rm TeV$ which corresponds to $R <
13\;\rm pm$. So the collider experiments reach or already supersede
the sensitivity of the short distance gravity experiments.

At the Tevatron direct graviton emission would manifest itself via the
production of single jets e.g. from gluons (monojets) along with large
missing transverse momentum. CDF and D{\O} have searched for such
topologies in the Run-I as well as in the Run-II data. In the absence
of a signal exclusion limits of about $M_S^{n=2} > 1\;\rm TeV$ have
been set. Table~\ref{tab:3} summarizes the searches for large extra
dimensions at the Tevatron in the various search channels. With the
continuing data taking of the Run-II further significant improvements
in sensitivity are expected.

\begin{table}[ht]
\caption{Summary of searches for large extra dimensions at the Tevatron
         along with resulting limits on fundamental scale in the ADD
         or RS model.}
\label{tab:3}       
\centerline{
\begin{tabular}{lrll}
\hline\noalign{\smallskip}
\multicolumn{2}{l}{experiment} & channel & limit\\
\noalign{\smallskip}\hline\noalign{\smallskip}
CDF-I &($110\;\rm pb^{-1}$)  & di-EM (e,$\gamma$) & $M_s > 0.94\;\rm TeV$\\
CDF-I &($87\;\rm pb^{-1}$)   & Mono-Jet/$\gamma$+$\not{\hspace{-1mm}E_T}$ & $M_s^{n=2} > 1.0\;\rm TeV$\\
CDF-I &($87\;\rm pb^{-1}$)   & Mono-Jet/$\gamma$+$\not{\hspace{-1mm}E_T}$ & $M_s^{n=6} > 0.6\;\rm TeV$\\
CDF-II &($75\;\rm pb^{-1}$)  & Di- e,$\mu, \gamma$, Jet) & $k/M_s$ limits in RS\\
\noalign{\smallskip}\hline\noalign{\smallskip}
D{\O}-I &($127\;\rm pb^{-1}$)  & Di-EM (e,$\gamma$) & $M_s > 1.2\;\rm TeV$\\
D{\O}-I &($78.8\;\rm pb^{-1}$) & Mono-Jet + $\not{\hspace{-1mm}E_T}$     & $M_s^{n=2} > 1.0\;\rm TeV$\\
D{\O}-I &($78.8\;\rm pb^{-1}$) & Mono-Jet + $\not{\hspace{-1mm}E_T}$     & $M_s^{n=6} > 0.65\;\rm TeV$\\
D{\O}-II &($120\;\rm pb^{-1}$) & Di-EM (e,$\gamma$) & $M_s > 1.28\;\rm TeV$\\
D{\O}-II &($30\;\rm pb^{-1}$)  & Di-$\mu$           & $M_s > 0.79\;\rm TeV$\\
\noalign{\smallskip}\hline
\end{tabular}}
\end{table}

\subsection{Search for Radions and Branons}
\label{sec:5.3}

In the framework of the Randall-Sundrum model massless and massive
spin-two excitations are predicted. The massless excitations couple
with gravitational strength and can be identified with gravitons.  The
masses and couplings of the massive spin-two excitations are set by
the weak scale. These states have not yet been observed and should, if
existent in nature, become visible with the next generation of
colliders. The spinless excitations, called {\it radions}, correspond
to a local fluctuation of the brane distance: $r_0 \rightarrow r_0 +
\Delta r(x)$. A mechanism has been proposed to stabilize the brane
distance, which is required to avoid the branes to drift apart faster
than compatible with cosmological models and observations. The radion
acquires a mass due to this stabilization mechanism, which is a free
parameter. The radion mass is expected to be well below $1\;\rm TeV$
and most likely lighter than massive spin-two excitations. If the
Randall-Sundrum model describes nature the radion is expected to be
the first sign of it which can be observed.

\begin{figure}[ht]
\resizebox{0.48\textwidth}{!}{%
  \includegraphics{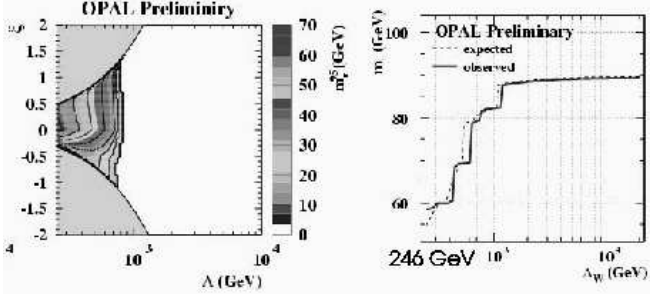}
}
\caption{Left: Observed radion mass limit as a function of the mixing 
         parameter $\xi$ and the fundamental scale $\Lambda_W$; Right:
         Lowest expected and observed limit on the Higgs boson mass as
         a function of the scale parameter $\Lambda_W$ for all allowed
         $\xi$ and for masses of the radion-like state $m_r$.}
\label{fig:12}       
\end{figure}

The radion carries the same quantum numbers as the Higgs boson, so
that these two particles can mix. A first search for radions has been
performed by OPAL \cite{opal_radion}, exploiting that the radion as
well as the SM Higgs boson are mainly produced in the Higgsstrahlung
process $e^+ e^- \rightarrow Zr$ or $\rightarrow Zh$. Limits on the
Higgsstrahlung cross section obtained from searches for the SM Higgs
boson, flavour-independent searches for hadronically decaying Higgs
bo\-sons and decay mode independent searches for Higgs bo\-sons are
used to restrict the parameter space of the RS model. As shown in
Figure~\ref{fig:12} the data excludes masses for the Higgs-like state
below $58\;\rm GeV$ for all scales $\Lambda_W \ge 246\;\rm GeV$
independent of the mixing between the radion and the Higgs boson, and
the radion mass $m_r$ in the range $1\;\rm MeV$ to $1\;\rm TeV$. The
analyses are sensitive to the radion only for scales $\Lambda_W \le
0.8\;\rm TeV$. No universal limits on the mass of the radion-like
state can be extracted as they depend on the mixing parameter $\xi$.

In the context of the ADD model new particle fields, called `branons'
are expected and associated with brane fluctuations along the extra
dimensions. The dynamics of these fields is determined via an
effective theory with couplings of order of the brane tension scale,
$f$. The search for Kaluza Klein gravitons and the search for branons
are complementary. If the brane tension is below the gravity scale, $f
\ll M_S$, the first signal of extra dimensions will come from branons,
allowing a measurement of $f$, the number of branons and their masses
$M$. If $f \gg M_s$, then the first evidence for extra dimensions will
be the discovery of gravitons, giving information about the
fundamental scale of gravitation ($M_s$) and the characteristics of
the extra-space (number of dimensions (D), size ($R_B$), topology
etc.).

\begin{figure}[ht]
\centerline{
\resizebox{0.30\textwidth}{!}{%
  \includegraphics{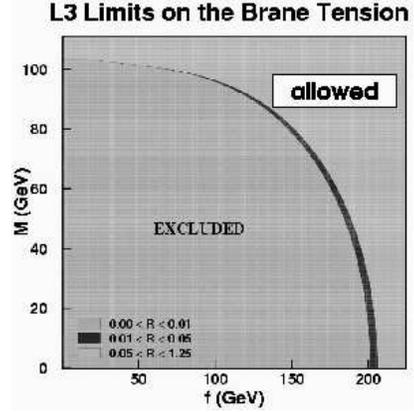} }
}
\caption{Two-dimensional limits in the $(f, M)$ plane from the analysis
         of L3 single photon data. For very elastic branes, $f
         \rightarrow 0$, the exclusion interval for the branon mass is
         $M \ge 103\;\rm GeV$.}
\label{fig:13}       
\end{figure}

Branons have a well defined effective theory with couplings to
Standard Model particles. Since processes involving the production of
only one branon are forbidden by Lorentz invariance the lowest order
branon interaction, which has been investigated by the L3
Collaboration \cite{l3_branons}, is $e^+ e^- \rightarrow \hat{\pi}
\hat{\pi} \gamma$, resulting in a single-photon final state. Comparing
the photon energy and angular distributions with the expectation of SM
or branon production limits have been placed in the $(f, M)$ plane
(see Figure~\ref{fig:13}) at 95\% CL ($R < 0.05$) and at 99\% CL ($R <
0.01$). Assuming a flexible brane ($f \ll M$) the single-$\gamma$
analysis with L3 data imposes the restriction: $f > 206\;\rm GeV$, a
similar single-$Z$ one: $f > 47\;\rm GeV$. Both limits are stronger
than the most stringent astrophysical constraint from the energy loss
observation in SN1987A; $f> \cal{O}$$(10)\;\rm GeV$.

Further details on searches for large extra dimensions, radions and
branons can be found in \cite{more_eps03_ed} and references therein.

\section{Searches for Supersymmetry}
\label{sec:6}
Supersymmetric (SUSY) extensions of the Standard Mo\-del, introducing
a symmetry between fermions and bosons, are particularly attractive as
they address various shortcomings of the Standard Model such as
quadratic divergences in the Higgs mass (cancel in SUSY), the fine
tuning problem, the hierarchy problem, coupling unification at very
large scales etc. As the supersymmetric partner particles of the known
Standard Model particles are not observed, SUSY must be broken.
Different models for the SUSY breaking mechanism have been considered,
in particular the minimal supersymmetric extension of the Standard
Model (MSSM), where the communication between the SUSY breaking sector
and our world is either established via gauge bosons (gauge mediated
SUSY breaking - GMSB with 6 parameters) or via gravity (gravity
mediated SUSY breaking - mSUGRA with 5 parameters). Also R-parity
violating SUSY models, where the SUSY particles are not necessarily
produced or decay in pairs, are considered. A very large number of
topologies has been predicted and searched for in the context of SUSY
in the recent years. Here only two examples are discussed: Searches
for stop and sbottom production and limits on chargino and neutralino
production. For further details, also on other SUSY searches, see
\cite{kuze_sirois,eps03_susy,eps03_rott}.

The states $\tilde{q_L}$ and $\tilde{q_R}$ are the partners of the
left-handed and the right-handed quarks. They form squark mass
ei\-gen\-states ($\tilde{q_1}$ and $\tilde{q_2}$), which are
orthogonal combinations of them. The mixing angle $\Theta_{\tilde{q}}$
is defined in such a way that $\tilde{q_1} = \tilde{q_L} \cos
(\Theta_{\tilde{q}}) + \tilde{q_R} \sin (\Theta_{\tilde{q}})$ is the
lighter squark. Since the off-diagonal elements of the mass matrix are
proportional to the mass of the corresponding SM partner, $m_q (A_q -
\mu \kappa)$, the mixing is expected to be relevant for the fermions
of the third family. In that expression $\mu$ is the Higgs mass
parameter, $A_q$ the trilinear coupling to the Higgs sector and
$\kappa = \tan \beta$ for down-type, $\kappa = 1 / \tan \beta$ for
up-type quarks. Therefore the sbottom quark could be light, if $\tan
\beta$ is large. A light stop could be realized due to large mass
splitting resulting from the large top mass. At hadron colliders stops
or sbottoms would be produced via the strong interaction, at LEP
electroweakly. The coupling of the stop and sbottom to the $Z$, and
therefore the production cross-section, are maximal for
$\Theta_{\tilde{q}} = 0$ and minimal for $\Theta_{\tilde{q}}= 56^0$
and $68^0$, respectively.

\begin{figure}[ht]
\centerline{
\resizebox{0.49\textwidth}{!}{%
  \includegraphics{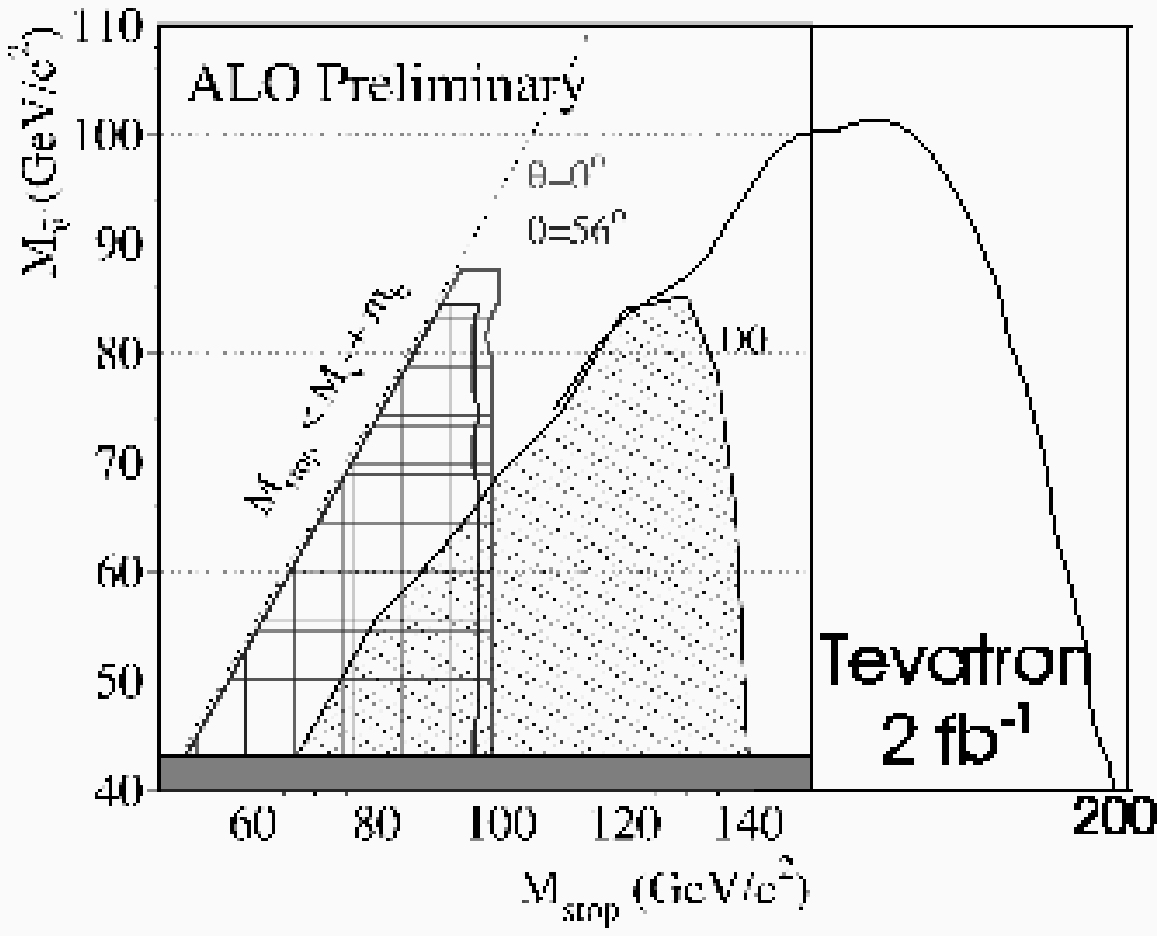} 
  \includegraphics{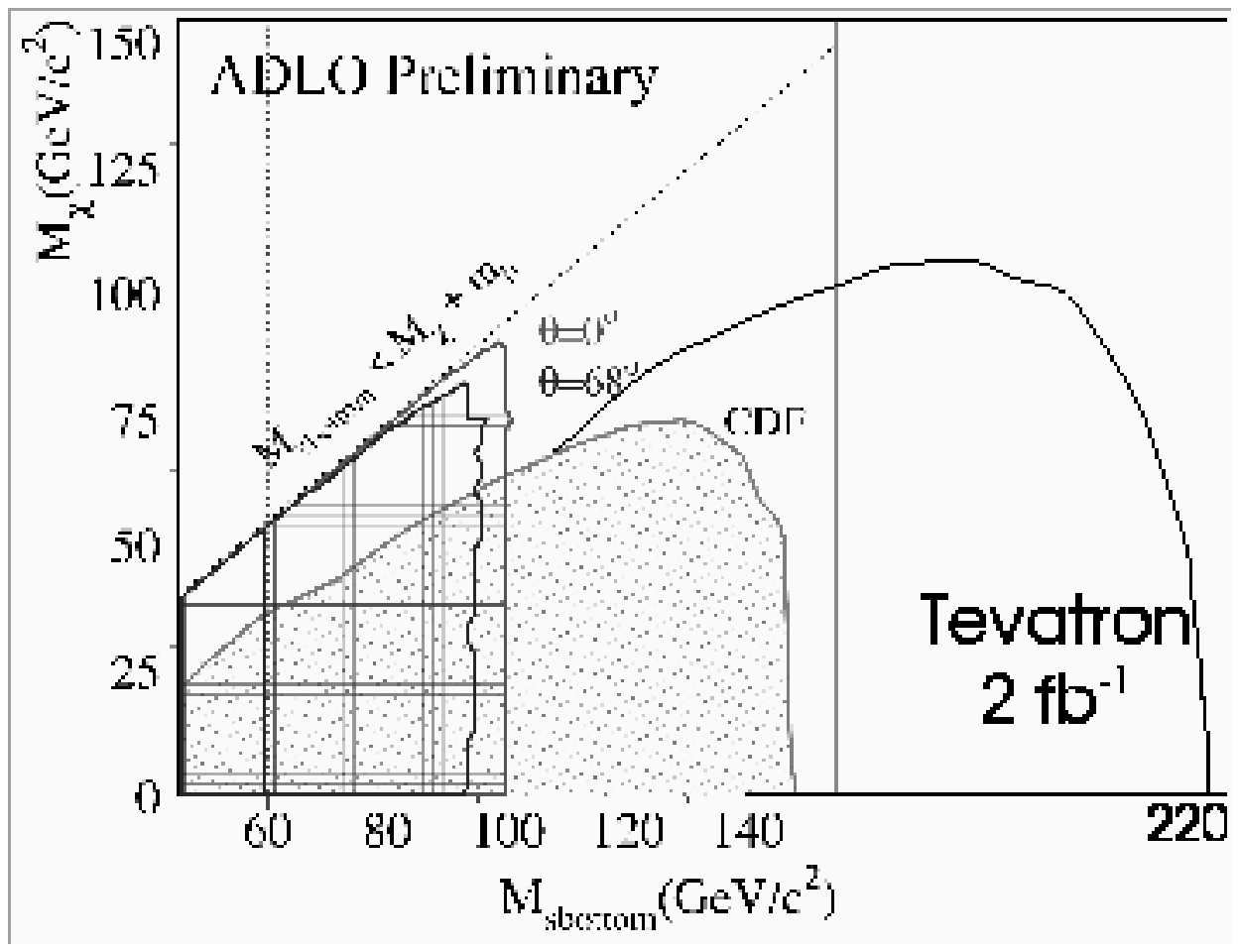} }}
\caption{Left: Stop and sneutrino mass
         plane showing the regions excluded by LEP, D{\O} and the
         expected sensitivity for $2\;\rm fb^{-1}$ at the Tevatron;
         right: Sbottom and neutralino mass plane showing the regions
         excluded by LEP, CDF and the expected sensitivity for $2\;\rm
         fb^{-1}$ at the Tevatron.}
\label{fig:14.1}       
\end{figure}

The decay $\tilde{t} \rightarrow t \tilde{\chi^0}$ is expected to be
dominant, but due to the large top mass and the limits on
$\tilde{\chi^0}$ currently experimentally not accessible. The
three-body decay $\tilde{t_1} \rightarrow b l \tilde{\nu}$ via
chargino would become the dominant decay mode if kinematically
allowed. Figure~\ref{fig:14.1} shows the LEP and D{\O} exclusions
limits in the ($m_{stop}, m_{\tilde{\nu}}$) plane. The LEP limits
reach up to $m_{stop} \approx 95\;\rm GeV$ and close to the kinematic
limit of $m_{stop} = m_{\tilde{\nu}} + m_b$, while the Tevatron reach
of up to $140\;\rm GeV$ in $m_{stop}$ can only be achieved with 20-30
GeV distance to the kinematic limit. The Tevatron experiments are
expected to reach up to $200\;\rm GeV$ in sensitivity with the first
$2\;\rm fb^{-1}$ of Run-II data. Similarly search limits of up to
$160\;\rm GeV$ have been achieved by the Tevatron experiments for the
loop suppressed flavour changing two body decay $\tilde{t_1}
\rightarrow c \tilde{\chi_1^0}$.
Sbottom quarks $\tilde{b_1}$ are expected to decay into $b
\tilde{\chi_1^0}$. The resulting exclusion limits in the 
($m_{stop}, m_{\tilde{\chi}}$) plane from LEP reaching close to $100
\;\rm GeV$ in $m_{sbottom}$, and CDF, reaching beyond $140\;\rm GeV$, 
are shown in Figure~\ref{fig:14.1}. The expected sensitivity of the
Tevatron experiments with the first $2\;\rm fb^{-1}$ reaches up to
$220\;\rm GeV$ in $m_{sbottom}$. Further sbottom quark searches from
gluino pair production with subsequent decay $\tilde{b_1} \rightarrow
b \tilde{\chi_1^0}$ are being performed by CDF-II and described in
\cite{eps03_rott}.

Searches for chargino pair production at LEP yielded mass exclusion
limits of $m_{\tilde{\chi_1^+}} \ge 103.5\;\rm GeV$ for
$m_{\tilde{\nu}} > 300\;\rm GeV$ which have been translated into mass
limits on the neutralino as a function of $\tan \beta$, assuming SU(5)
and SO(10) GUT-relations. In the constrained MSSM, where the
neutralino is the lightest SUSY particle (LSP), an absolute mass limit
of $m_{LSP} > 46\;\rm GeV$ is found \cite{lep_neutralino_limit}. If,
however, these GUT relations are dropped, assuming for example
unification via string theory, no bounds on the neutralino mass can be
placed anymore by collider experiments. Assuming, however, the LSP to
be the lightest neutralino, to be responsible for the observed cold
dark matter relic density and from respecting the LEP2 limits on
chargino, slepton and sneutrino masses \cite{lsp_cdm}, still a limit
of $m_{LSP} > 5\;\rm GeV$ is obtained. Dropping those assumptions as
well, only requiring a possible LSP to be consistent with the observed
time dependence of the signal from the supernovae SN1987A, still a
limit on the neutralino mass of $m_{LSP} > 100\;\rm MeV$ can be placed
\cite{lsp_sn1987}.

New collider results are expected soon from the di- and tri-lepton
searches at the Tevatron Run-II.

For the first time at a hadron collider, the CDF and D{\O}
Collaborations have observed $Z \rightarrow \tau \tau$ signals, where
one $\tau$ decays leptonically and the other $\tau$ is identified in
hadronic one- or three-prong decays. This progress in
$\tau$-identification, using sophisticated experimental methods,
represents a milestone in SUSY and Higgs searches as it opens a
sensitivity window to many models and searches for particles coupling
to the third fermion generation.


\section{Standard Model Higgs Boson Searches}
\label{sec:7}

The Higgs mechanism \cite{p_higgs} plays a central role in the
unification of the electromagnetic and weak interactions by providing
mass to the $W$ and $Z$ intermediate vector bosons without violating
local gauge invariance. Within the Standard Model, the Higgs mechanism
is invoked to break the electroweak symmetry; it requires one doublet
of complex scalar fields which leads to the production of a single
neutral scalar particle, the Higgs boson. The mass of this particle is
not specified, but indirect experimental limits are obtained from
precision measurements of the electroweak parameters which depend
logarithmically on the Higgs boson mass through radiative
corrections. Currently these measurements predict the Standard Model
Higgs boson mass to be $m_H = 91 ^{+ 58}_{- 37}\;\rm GeV$ and
constrain its value to less than $219\;\rm GeV$ at the 95\% confidence
level
\cite{eps_pippa,lepew_hfit}.

\subsection{SM Higgs Boson Searches at LEP}

The data collected by the four LEP collaborations prior to the year
2000 gave no direct indication of the production of the Standard Model
Higgs boson and allowed a lower bound of $107.9\;\rm GeV$ to be set on
the mass. During the last year of the LEP programme (the year 2000),
substantial data samples were collected at centre-of-mass energies
$(\sqrt{s})$ exceeding $206\;\rm GeV$, extending the search
sensitivity to Higgs boson masses of about $115\;\rm GeV$ through the
Higgs\-strahlung process $e^+e^- \rightarrow HZ$. In their initial
ana\-lyses of the full data sets, ALEPH observed an excess of events
consistent with the production of a Standard Model Higgs boson with a
mass of $115\;\rm GeV$; L3 and OPAL, while being consistent with the
background hypothesis, slightly favoured the signal plus background
hypothesis in this mass region; DELPHI reported a slight deficit with
respect to the background expectation.

The final results from the four collaborations have by now been
published. These are based on full data reprocessing using final
calibrations of the detectors and LEP beam energies, new and improved
Monte Carlo generators and increased Monte Carlo event statistics, in
some cases, on revised analysis procedures. In total a data set of
$2461\;\rm pb^{-1}$ at $\sqrt{s} > 189\;\rm GeV$ ($536\;\rm pb^{-1}$
at $\sqrt{s} > 206\;\rm GeV$) has been analysed and combined by the
LEP-Higgs working group using the likelihood ratio technique
\cite{leph_smh_final}.

\begin{figure}[ht]
\resizebox{0.49\textwidth}{!}{%
  \includegraphics{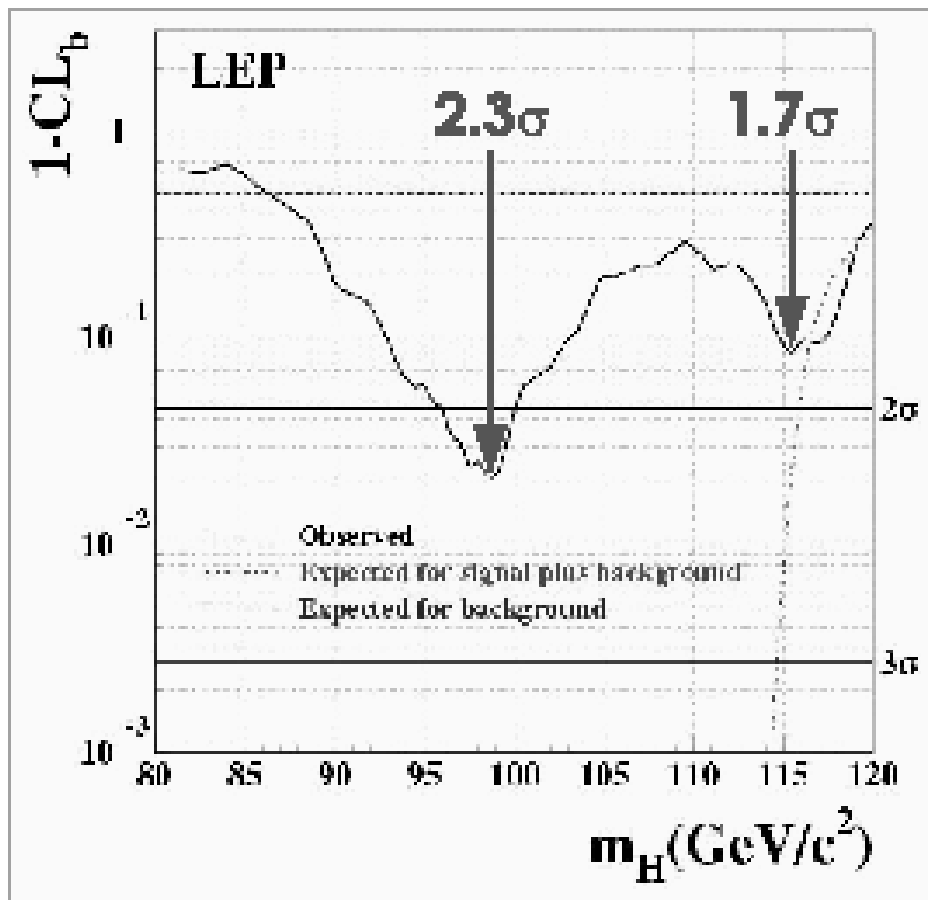}
  \includegraphics{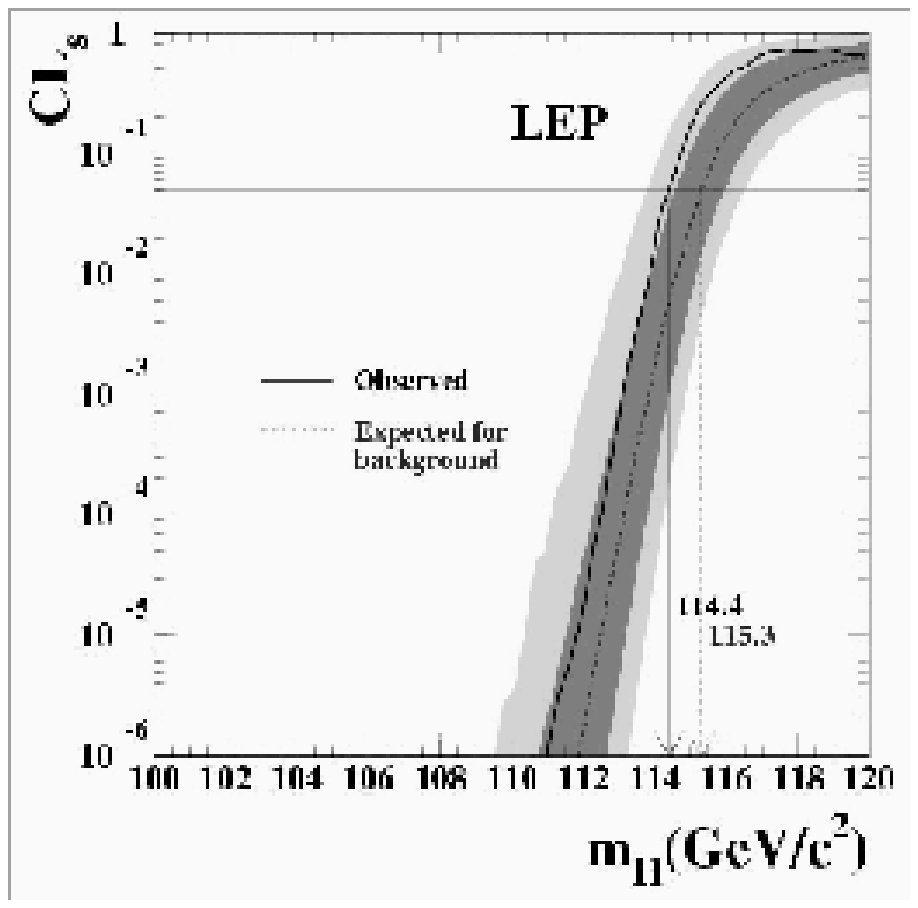}
}
\caption{Left: The background confidence level $1 - CL_b$ as a function 
         of the test mass $m_H$. The full curve is the observation;
         the dashed curve is the median expected background confidence
         and the dashed dotted line is the median expectation for $1 -
         CL_b$, given the signal plus background hypothesis, when the
         signal masses on the abscissa is tested. Right: The ratio
         $CL_s = CL_{s+b} / CL_b$ for the signal plus background
         hypothesis, as a function of the test mass $m_h$. The solid
         line is the observation and the dashed line the median
         background expectation. The dark and light shaded bands
         around the median expectation for the background hypothesis
         correspond to the 68\% and 95\% probability bands.}
\label{fig:15}       
\end{figure}

In Figure~\ref{fig:15} the final results are summarized. Combining the
results from the four LEP experiments a lower bound of $114.4\;\rm
GeV$ is set on the mass of the Standard Model Higgs boson. The mass
limits from the single experiments and the combination are summarized
in Table~\ref{tab:4}. At a mass of $115\;\rm GeV$, where ALEPH
reported an excess compatible with the production of a Standard Model
Higgs boson, the confidence $1 - CL_b$ of the combined LEP data
expressing the level of consistency with the background hypothesis is
0.9 ($\sim 1.7 \sigma$), while the confidence level $CL_{s+b}$
measuring the consistency with the signal plus background hypothesis
is 0.15. In the region $m_H \approx 98\;\rm GeV$ the observed value of
$1 - CL_b \approx 0.02$ translates into 2.3 standard deviations. This
excess, however, is clearly no sign of Higgs boson production since
the number of events expected for such a signal would be a factor of
ten larger than the one observed.

\begin{table}[ht]
\caption{Final lower Higgs mass limits of the four LEP-experiments 
         and their combination.}
\label{tab:4}       
\centerline{
\begin{tabular}{lll}
\hline\noalign{\smallskip}
experiment & expected $M_h >$ & observed $M_h >$ \\
\noalign{\smallskip}\hline\noalign{\smallskip}
ALEPH  & 113.5 GeV & 111.5 GeV\\
DELPHI & 113.3 GeV & 114.2 GeV\\
L3     & 112.4 GeV & 112.0 GeV\\
OPAL   & 112.7 GeV & 112.8 GeV\\ 
\noalign{\smallskip}\hline
LEP    & 115.3 GeV & 114.4 GeV\\
\noalign{\smallskip}\hline
\end{tabular}}
\end{table}

The searches for the Standard Model Higgs boson carried out by the
four LEP experiments extended the sensitive range well beyond that
anticipated at the beginning of the LEP programme. This is due to the
higher energy achieved and to more sophisticated detectors and
analysis techniques.

\begin{figure}[ht]

\centerline{
\resizebox{0.41\textwidth}{!}{%
  \includegraphics{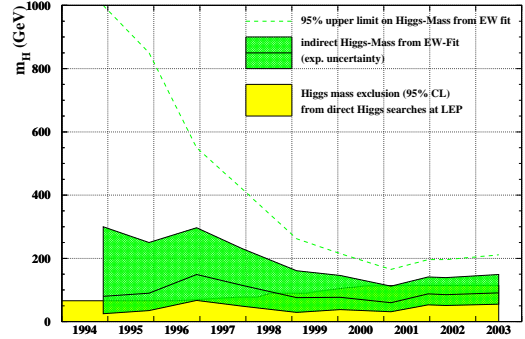}
}}
\caption{Development of the upper higgs mass limit (dashed line) and the
         central value and uncertainty of the electroweak fit (hashed region)
         over the last years. The light shaded region shows the higgs masses
         which have been excluded by direct searches for the Higgs boson.}
\label{fig:16}       
\end{figure}

Figure~\ref{fig:16} shows the development of the direct and the
indirect mass bounds over the last years. While the upper mass limit
(95\% CL) decreased from $1000\;\rm GeV$ to about $200\;\rm GeV$ the
best fit value oscillated between $60$ and about $150\;\rm GeV$ with a
strongly decreasing uncertainty of the electroweak fit. The Higgs
boson mass limit from the direct searches at LEP has been steadily
increasing from $66$ to $114.4\;\rm GeV/c^2$.  The future searches for
the Standard Model Higgs boson will concentrate at the low mass region
around $110 - 140\;\rm GeV$, where the Tevatron experiments have the
sensitivity to discover or exclude the Higgs boson before the LHC
turns on.

\subsection{SM Higgs Boson Searches at the Tevatron}

Already the preliminary results on the search for the Standard Model
Higgs boson at LEP suggested that the Higgs boson mass may not be very
high, which inspired corresponding studies at the Tevatron for Run-II.

The Higgs boson production mechanism with the lar\-gest cross section,
$\sim 0.7\;\rm pb$ for a Higgs mass of $120\;\rm GeV$, is the gluon
fusion. Unfortunately the background in this mode is very large. The
most promising Standard Model Higgs discovery channels at the Tevatron
are the associated production with $W/Z$, where the $W/Z$ decay
leptonically. The Higgs boson production cross section is $\sim
0.16\;\rm pb$ for $WH$ and $\sim 0.1\;\rm pb$ for $ZH$ production of a
$120\;\rm GeV$ Higgs boson. 


Using Run-I data, CDF has searched for $WH$ and $ZH$ in different
channels, including $Z$ decaying to dileptons, $W$ decaying to leptons
and hadrons and the Higgs decaying to $b\bar{b}$. The results from
combining these channels for a $130\;\rm GeV$ Higgs mass gives
\begin{eqnarray}
\sigma(p\bar{p} \rightarrow VH) * Br(H \rightarrow b\bar{b}) & < & 7.4\;\rm pb\,\; at \,\; 95\% CL
\end{eqnarray}
A few years ago, a Tevatron Higgs Working Group's study evaluated the
Higgs discovery potential for the Tevatron Run-II
\cite{susyhiggs_runii_study}. This was a joint effort of theorists and
experimentalists from both, the CDF and D{\O} experiment. The study
was based on a parameterized detector simulation. The main conclusions
were that in order to maximize the Higgs discovery potential at the
Tevatron, one must combine data from both experiments, CDF and D{\O};
must combine all channels, and must improve the understanding of
signal and background processes as well as improve the detector
performance. About $2\;\rm fb^{-1}$ of data per experiment were
expected to be required for a 95\% CL exclusion of a $115\;\rm GeV$
mass Higgs boson, while a $3 \sigma$ observation was expected to be
possible with $5 - 6\;\rm fb^{-1}$ and a $5 \sigma$ discovery with
$15\;\rm fb^{-1}$ per experiment. The Tevatron experiments were at the
time of this study (1999) expected each to collect $2\;\rm fb^{-1}$ in
Run-IIA and $15\;\rm fb^{-1}$ in Run-IIB, i.e. by 2006, so that a
discovery would have been possible for a $\sim 115\;\rm GeV$ mass
Higgs boson and in the absence of a Higgs boson an exclusion potential
of up to $\sim 180\;\rm GeV$ was estimated.

In the meantime CDF and D{\O} reviewed and repeated the Higgs
sensitivity study \cite{higgs_sensitivity_ii}, taking the projections
of the already achieved Run-II performance of the largely and
successfully upgraded detectors and the improved understanding of the
dominant background sources from data measurements into
account. Particular emphasize was gi\-ven to the calorimeter
resolutions on jet-$E_T$ and di-jet mass resolutions as well as the
tracking impact parameter resolution, one of the crucial detector
parameters for the $b$-tagging algorithms. The expectations of the
initial study \cite{susyhiggs_runii_study} are all essentially met,
while the analyses techniques and background estimates could be
significantly improved.  The studies were split up such that CDF
reviewed the $WH \rightarrow l\nu b\bar{b}$ channel while D{\O}
reviewed the $ZH \rightarrow \nu \nu b\bar{b}$ channel. Assuming
comparable performance by CDF and D{\O} the results were then
combined. In the following, the $ZH \rightarrow \nu \nu b\bar{b}$
analysis is review in more for detail as an example.

\begin{table*}[ht]
\caption{Expected number of events per $fb^{-1}$ data in the 
         $ZH \rightarrow \nu \nu b\bar{b}$ analysis as estimated at
         the SUSY/Higgs Workshop in 1999 (SHW 1999), as estimated
         using the new neural network analysis, but identical cross
         sections (ANN'03) and as estimated using that new analysis
         with the latest cross section calculations or estimates from
         data (Analysis'03).}
\label{tab:5}       
\centerline{
\begin{tabular}{|c||r|r|r|r|r||l|}
\hline\noalign{\smallskip}
Process & SHW 1999 & ANN'03 & Ratio & Analysis'03 & Ratio & Comment\\
\noalign{\smallskip}\hline\noalign{\smallskip}
$HZ (115\;\rm GeV)$ & 3.15 & 3.82 & 1.22 & 2.86 & 0.91 & \\
$HW (115\;\rm GeV)$ & 2.39 & 2.78 & 1.16 & 2.08 & 0.87 & \\
\noalign{\smallskip}\hline\noalign{\smallskip}
$Zb\bar{b}$  & 4.34 & 1.73 & 0.4\phantom{0} & 1.99 & 0.46 & estimated from CDF data\\
$Wb\bar{b}$  & 9.45 & 3.59 & 0.38 & 4.34 & 0.46 & estimated from CDF data\\
$ZZ$         & 1.82 & 2.36 & 1.3\phantom{0}  & 2.93 & 1.61 & PYTHIA 6.125 + K-factor=1.34\\
$WZ$         & 1.45 & 1.79 & 1.45 & 1.84 & 1.27 & PYTHIA 6.125 + K-factor=1.34\\
$t\bar{t}$   & 3\phantom{.00}   & 6.53 & 2.18 & 5.48 & 1.83 & average of NLO calculations \\
$qtb$        & 0.31 & 0.8\phantom{0} & 2.62 & 0.68 & 2.22 & NLO calculation\\
$tb$         & 4.7\phantom{0}  & 0.49 & 0.1\phantom{0} & 0.35 & 0.08 & NLO calculation\\
QCD          & 25.06& 17.3 & 0.69 & 11.16& 0.45 & from this study\\
\noalign{\smallskip}\hline\noalign{\smallskip}
total background & 50.11 & 34.59 & & 28.77 & & \\
significance $(S / \sqrt{B})$    & 0.78  & 1.12  & & 0.92  & &  \\
\noalign{\smallskip}\hline
\end{tabular}}
\end{table*}

This new Higgs sensitivity study was done in two stages: In the first
stage the analyses assumptions were left unchanged with respect to the
previous study in \cite{susyhiggs_runii_study} while the analysis was
improved in a number of aspects. In particular:
\begin{itemize}
\item The QCD background level was conservatively estimated to be 
      equal to the level of the remaining background sources,
      i.e. 50\% of the total background was assumed to originate from
      QCD processes.
\item The trigger efficiency was assumed to be 100\%.
\item The $b$-tagging efficiency was assumed to be 35\% for events 
      with one tight and one loose tag, and 32\% for events with two
      tight tags.
\item In contrast to \cite{susyhiggs_runii_study}, where the detector
      performance was estimated using fast parametrisations, here it
      was simulated using the full GE\-ANT detector description.
\item The analysis was based on a real artificial neural network, while
      previously it was cut-based.
\item The Monte Carlo event samples in the new study were obtained by
      mixing the hard collision events with five additional minimum
      bias events overlayed while previously only hard scattering
      events were considered.
\end{itemize}

The number of events selected per $fb^{-1}$ in the $ZH \rightarrow \nu
\nu bb$ analysis for a Higgs of mass $115\;\rm GeV$ is summarized in
Table~\ref{tab:5}. It can clearly be seen that the main change comes
from the large reduction in the $Zbb$ and $Wbb$ events, exploiting the
separation power of the neural network, resulting in a significant
reduction of the total expected background. This translates into a
significance increase, estimated in $S/\sqrt{B}$, from 0.78 to 1.12,
where the Higgs signal mostly originates from the $HZ$ production but
also $HW$ significantly contributes to the $HZ \rightarrow bb \nu \nu$
analysis. In other words the integrated luminosity, required for a
particular significance, has been reduced by $\sim 50\%$.

In a second stage new cross section estimates, either from improved
theory calculations or from measurements using Tevatron data (here
upper limits on the cross sections for the $Wbb$ and $Zbb$ process)
and an improved description of the analyses was applied. In
particular:
\begin{itemize}
\item Estimates of realistic trigger efficiencies were applied.
\item The QCD background contribution was estimated from data.
\item The $b$-tagging efficiency was assumed to be 35\% for events 
      with one tight and one loose tag, and 32\% for events with two
      tight tags.
\end{itemize}

From the summary in Table~\ref{tab:5} it can be seen that the largest
improvement comes from the reduction in QCD background, which is to a
large extent instrumental background and best estimated from the data
itself. This more realistic study results in a significance increase
from 0.78 to 0.92, which translates into the integrated luminosity,
required for a particular significance, still being reduced by $\sim
28\%$.
Given those numbers it should be noticed that:
\begin{itemize}
\item At present the double $b$-tagging efficiency for Run-IIA data
      has been found to be 19\% compared to the 32\% expected from the
      upgrade to the Run-IIB silicon vertex detectors, which is
      assumed for this as well as the previous study. If the vertex
      detectors were not upgraded, more luminosity would be
      needed\footnote{In September 2003 it was decided by the Fermilab
      Directorate to cancel the Run-IIB silicon vertex upgrade
      programmes. Consequently the required luminosity for a given
      Higgs sensitivity is larger and needs to be re-evaluated.}
\item More sophisticated statistical combination techniques as used 
      in the LEP-Higgs search ($CL_s$ method), using mass
      distributions rather than simple event counting as used here,
      are expected to reduce the required luminosity for a given
      sensitivity by $\sim 20\%$.
\item More sophisticated analysis techniques are being developed, 
      which are expected to reduce the required data statistics for a
      given sensitivity even further. See the latest Run-I top mass
      reanalysis for example \cite{runi_top_mass}.
\item This analysis is still not optimized. Reviewing the 
      $H \rightarrow WW$ channels is expected to improve the
      sensitivity in a similar manner for Higgs boson masses at or
      above $130\;\rm GeV$.
\end{itemize}

\begin{figure}[ht]
\centerline{
\resizebox{0.49\textwidth}{!}{%
  \includegraphics{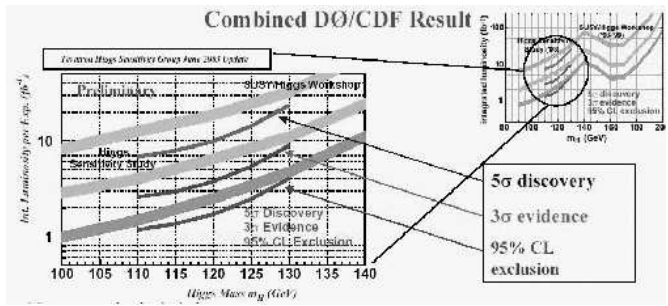} }
}
\caption{Integrated luminosity required per experiment, to either 
         exclude a SM Higgs boson at 95\% CL or to discover it at the
         $3\, \sigma$ or the $5\,\sigma$ level at the Tevatron.}
\label{fig:17}       
\end{figure}

Figure~\ref{fig:17} shows the result of the new Higgs sensitivity
study, combining the CDF and D{\O} results on the $WH \rightarrow l\nu
b\bar{b}$ and $ZH \rightarrow \nu \nu b\bar{b}$ channels,
respectively, and assuming equivalent improvements in the $ZH
\rightarrow ll bb$ channel. At this stage the combination was done
using the $CL_s$ method. The largely ($\sim 28\%$) reduced integrated
luminosity required for the 95\% CL exclusion, the $3\sigma$
observation, or the $5 \sigma$ discovery for a given Higgs boson mass
is shown as a function of the Higgs boson mass. Since the analy\-ses
concentrated on the decay $H \rightarrow b\bar{b}$, improvements with
respect to \cite{susyhiggs_runii_study} are seen in the Higgs boson
mass range up to $\sim 130\;\rm GeV$. Similar potential for
improvement at higher masses is expected from a review of the $H
\rightarrow WW$ decay mode channels. Since this study did not include
extensive systematic studies the wide error band, known from the
previous study, is not shown.

\begin{table}[ht]
\caption{Latest projections on integrated luminosity in $fb^{-1}$
         for the Tevatron Run-II in the conservative baseline and in
         the design scenario.}
\label{tab:6}       
\centerline{
\begin{tabular}{rrr}
\hline\noalign{\smallskip}
year & baseline & design \\
\noalign{\smallskip}\hline\noalign{\smallskip}
2003 & 0.28 & 0.3\\
2004 & 0.59 & 0.68\\
2005 & 0.98 & 1.36\\
2006 & 1.48 & 2.24\\
2007 & 2.11 & 3.78\\
2008 & 3.25 & 6.15\\
2009 & 4.41 & 8.57\\
\noalign{\smallskip}\hline
\end{tabular}}
\end{table}

Along with this reviewed Higgs sensitivity study by the CDF and D{\O}
collaborations, also the Tevatron luminosity performance and
projections have been reviewed. Based on the observed Tevatron
collider performance compared to the initial luminosity plans, new
projections for the integrated luminosity in Run-II have been worked
out in a conservative `baseline' and in the `design' scenario. These
numbers include a shutdown in 2005/2006 for installation of new
machine components and optimization and are summarized in
Table~\ref{tab:6}. Depending on the future development of the Tevatron
the collider experiments are expected to receive a delivered
integrated luminosity between $4.4\;\rm fb^{-1}$ and $8.6\;\rm
fb^{-1}$. These projections in lumi\-no\-sity compensate the improved
analysis sensitivity so that overall the Higgs mass region to which
the Tevatron experiments are sensitive for exclusion, observation or
discovery is somewhat reduced compared to the initial estimate
\cite{susyhiggs_runii_study}.

With the presently available data set the Tevatron experiments have no
sensitivity to the Standard Model Higgs boson. Nevertheless, searches
for the Higgs boson in the $HZ$ and $HW$ channels are being performed
and studied, in particular in the $H \rightarrow WW$ channel. A
detailed description of the present status is given in
\cite{eps03_varelas}.

\subsection{Higgs Boson Searches at the LHC}
The search for the Standard Model Higgs boson is one of the primary
tasks of the experiments at the {\it Large Hadron Collider} (LHC). It
has been established in many studies \cite{lhc_sm_higgs} that a
Standard Model Higgs boson can be discovered with high significance
over the full mass range of interest, from the lower limit of
$114.4\;\rm GeV$ set by the LEP experiments \cite{leph_smh_final} up
to about $1\;\rm TeV$.

At the LHC, the production cross-section for a Standard Model Higgs
boson is dominated by gluon-gluon fusion. The second largest
cross-section comes from the fusion of vector bosons radiated from
initial-state quarks. The relative contributions of the two processes
depend on the Higgs boson mass. For $m_H < 2 m_Z$, vector boson fusion
amounts to about 20\% of the total production cross-section and
becomes more important with increasing mass. However, for this
production mode, additional event characteristics can be exploited to
suppress the large backgrounds. The Higgs boson is accompanied by two
jets in the forward regions of the detector, originating from the
initial quarks that emit the vector bosons. In addition, central jet
activity is suppressed due to a lack of colour exchange between the
quarks. This is in contrast to most background processes, where there
is colour flow in the $t$-channel. Therefore jet tagging in the
forward region of the detector together with a veto of jet activity in
the central region are useful tools to enhance the
signal-to-background ratio.

The observation of the Standard Model Higgs boson at the LHC in the
vector boson fusion channels in the intermediate mass range was first
discussed in \cite{wbf_higgs_zepp1} for the $H \rightarrow \gamma
\gamma$ and $H \rightarrow WW^{(*)}$ decay modes and in 
\cite{wbf_higgs_zepp2} for the $H \rightarrow \tau \tau$ decay mode.
In the recent past ATLAS \cite{atlas_wbf_higgs} and CMS
\cite{cms_wbf_higgs} have performed analyses for the $WW^{(*)}$ and
$\tau \tau$ decay modes using realistic detector simulations of the
expected performance of the detectors at the LHC, including forward
jet tagging. The performance is addressed at low luminosity,
i.e. $\cal{L}$$= 10^{33}\;\rm cm^{-2} s^{-1}$. The discovery
potential is evaluted for an integrated luminosity up to $30\;\rm
fb^{-1}$, which is expected to be reached during the first few years
of operation.

\begin{figure}[ht]
\centerline{
\resizebox{0.49\textwidth}{!}{%
  \includegraphics{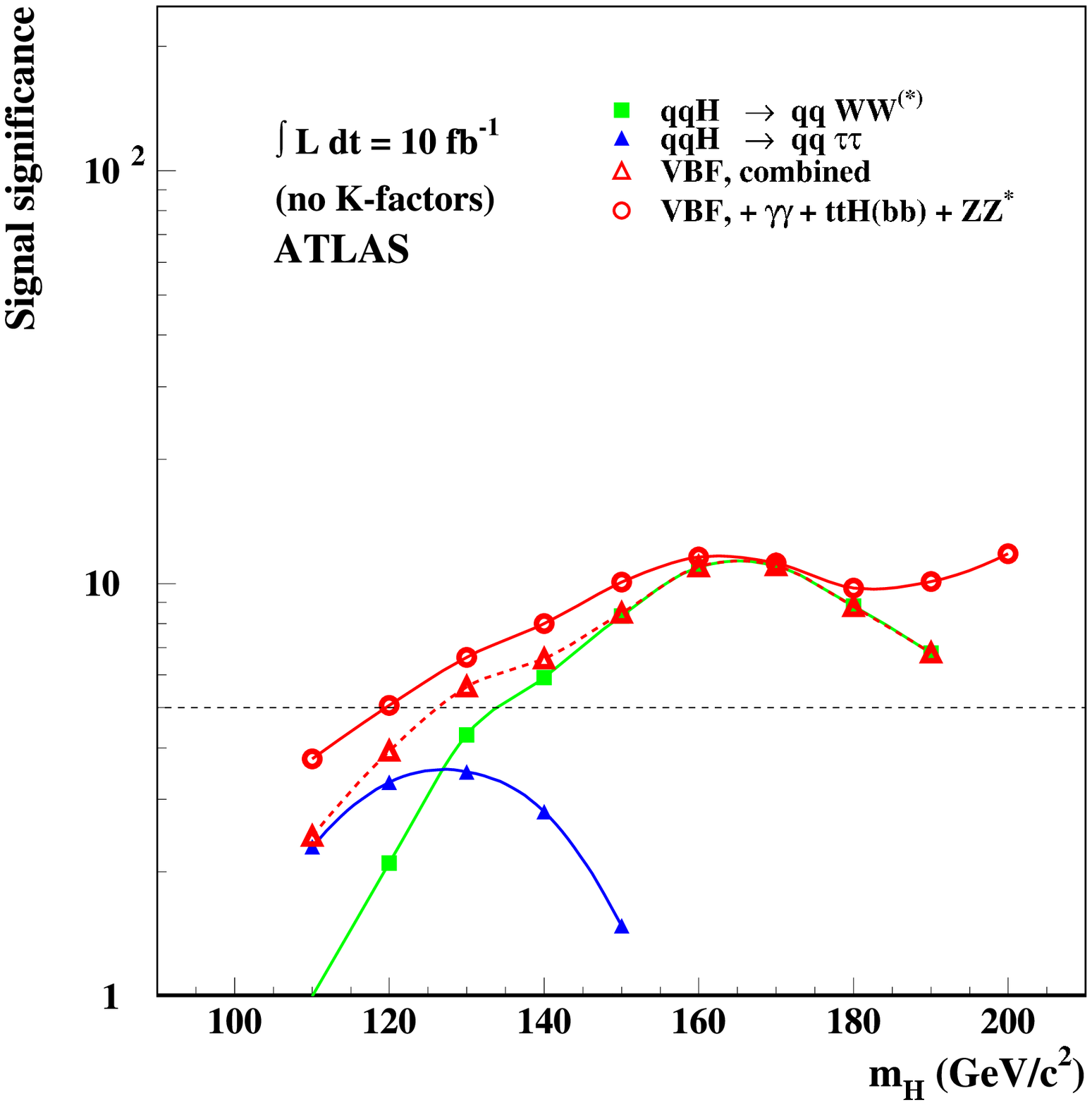} 
  \includegraphics{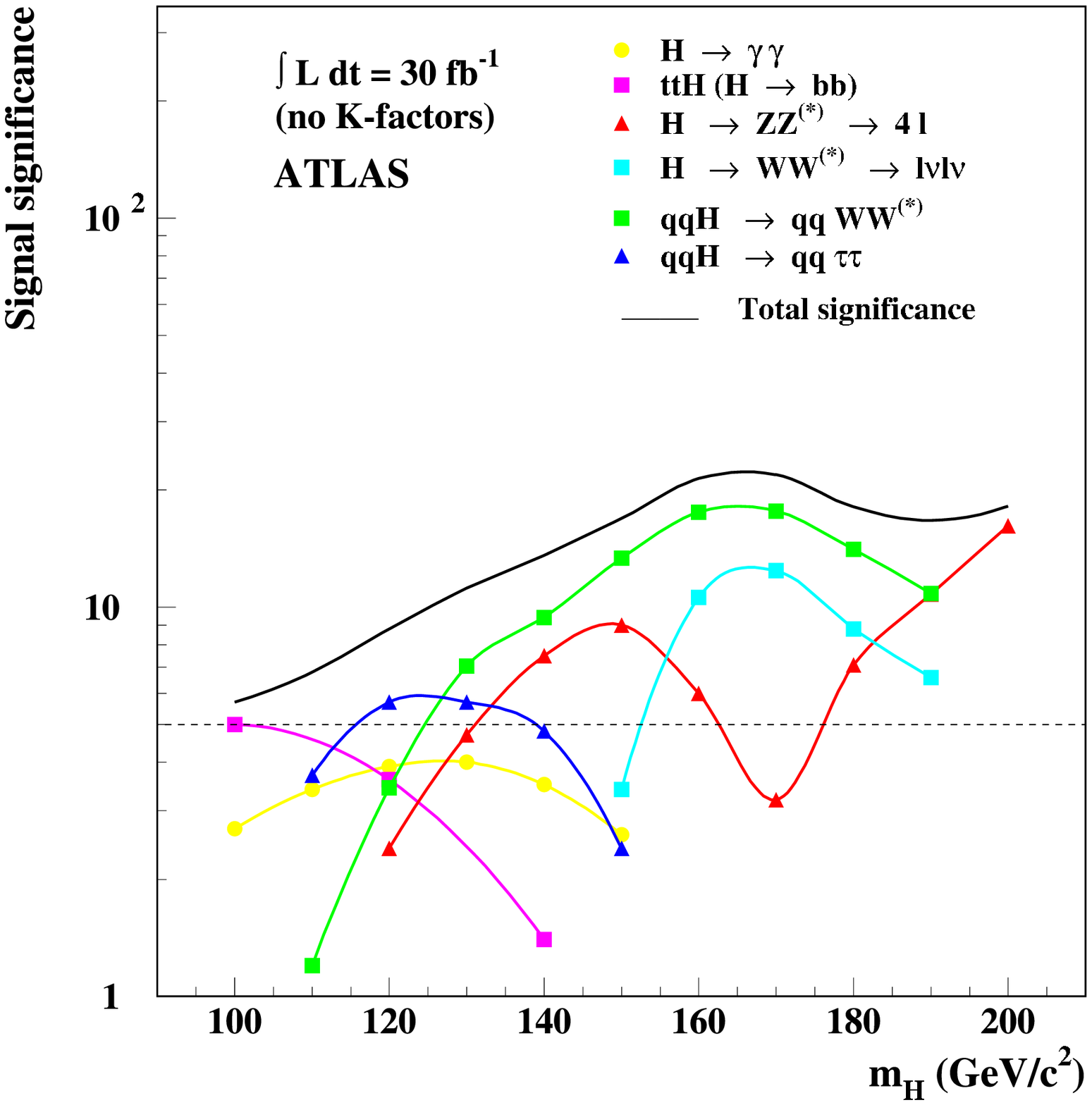} 
}
}
\caption{ATLAS sensitivity for the discovery of a Standard Model Higgs
         boson for integrated luminosities of $10$ and $30\;\rm
         fb^{-1}$. The signal significances are plotted for individual
         channels, as well as for the combination of channels.}
\label{fig:18}       
\end{figure}

In the example of the ATLAS experiment, the resulting sensitivity for
a discovery of the Standard Model Higgs boson in the mass range $110 -
190\;\rm GeV$ including the vector boson fusion channels is shown in
Figure~\ref{fig:18} for integrated luminosities of $10$ and $30\;\rm
fb^{-1}$. The vector boson fusion channels provide a large discovery
potential even for small integrated luminosities. Combining the two
vector boson fusion channels, a Standard Model Higgs boson can be
discovered with a significance above $5 \sigma$ in the mass range
$135$ to $190\;\rm GeV$ assuming an integrated luminosity of $10\;\rm
fb^{-1}$ and a systematic uncertainty of $\pm 10\%$ on the background.
If the vector boson fusion channels are combined with the standard
Higgs boson discovery channels $H \rightarrow \gamma \gamma$, $H
\rightarrow ZZ^{(*)} \rightarrow 4l$, and $t\bar{t}H$ with $H
\rightarrow b\bar{b}$, the $5 \sigma$ discovery range can be extended
down to $\sim 120 \;\rm GeV$.

For an integrated luminosity of $30\;\rm fb^{-1}$, the full mass range
can be covered by ATLAS with a significance exceeding $5 \sigma$. Over
the full mass range several channels, complementary both in physics
and detector aspects, will be available for a Higgs boson
discovery. The three different channels test different production
mechanisms, the gluon-gluon fusion via the $\gamma \gamma$ channel,
the vector boson fusion via the $WW$ and $\tau \tau$ channels, and the
associated $t\bar{t}H$ production in the $H \rightarrow b\bar{b}$
mode. This complementarity also provides sensitivity to non-standard
Higgs models, such as fermiophobic models, and allows a measurement of
the Higgs boson couplings to fermions \cite{eps03_duersen}. Similar
results have been obtained by the CMS Collaboration
\cite{cms_wbf_higgs}.

In the recent years also on the theoretical understanding of signal
and background calculations for the Higgs boson searches at the LHC a
lot of progress has been made, in particular on NLO calculations of
$pp / p\bar{p} \rightarrow t\bar{t}H$ \cite{nlo_tth}, NNLO
calculations on the inclusive Higgs production via $gg$-fusion
\cite{nnlo_gg_fusion}, NLO calculations on background processes for 
$gg \rightarrow \gamma \gamma$ \cite{nlo_gg_gaga}, NLO calculations on
$bg \rightarrow bH$ \cite{nlo_bh} and $WH$- and $ZH$-production
\cite{wh_zh}.


\section{Neutral Higgs Boson searches in the MSSM}
Numerous theoretical attempts to extend the SM usually lead to the
introduction of at least one additional doublet of scalar fields in
the Higgs sector. In the mi\-ni\-mal supersymmetric extension of the
Standard Model (MSSM) this results in 5 Higgs boson mass eigenstates -
3 neutral ($h / H / A$) and 2 charged ($H^+$ and $H^-$). In the
CP-conserving MSSM, the three neutral Higgs bosons are CP eigenstates:
the $h$ and $H$ are CP-even and the $A$ boson is CP-odd since all
CP-violating phases in the Higgs sector have been set to zero. Only
the CP-even states couple to the Z boson at tree level. Therefore the
main Higgs production mechanisms at LEP are the Higgsstrahlung $e^+e^-
\rightarrow hZ$ and $HZ$, and the associated production $e^+e^-
\rightarrow hA$ and $HA$.  Instead of one parameter - the mass of the
Higgs boson - the phenomenology is now described by two parameters
with the most popular choice of $\tan \beta$ and the mass of the
pseudoscalar neutral bosons $M_A$. At tree level the mass of the
lighter CP-even Higgs boson is restricted to be less than the mass of
the $Z^0$. Radiative corrections, in particular from loops containing
the top quark, allow the lightest Higgs boson mass to range up to
approximately $135\;\rm GeV$. The combined data of the four LEP
experiments are interpreted in the framework of the `constrained'
MSSM, where in the past three typical benchmark MSSM scans were
considered, yielding typical observed (expected) mass limits of $m_h >
91 \, (95)\;\rm GeV$ and $m_A > 92 \, (95) \;\rm GeV$ with $\tan
\beta$ exclusions between 0.5-2.4 or 0.7 - 10.5, depending on the
benchmark scan. Details on these searches can be found in
\cite{leph_mssm_cern}.

In the last years more MSSM benchmark scans have been proposed
\cite{lhc_scans}, inspired by Higgs Boson searches at the LHC.
Due to the different initial states, the Higgs production and decay
channels relevant for Higgs boson searches were different at LEP2 to
what they are at hadron colliders. New benchmark scenarios for the
MSSM Higgs boson search at hadron colliders have been suggested that
exemplify the phenomenology of different parts of the MS\-SM parameter
space. Besides the $m_h^{max}$ scenario and the no-mixing scenario
used in the LEP2 Higgs boson searches, two new scenarios have been
proposed. In one the main production channel at the LHC, $gg
\rightarrow h$, is suppressed. In the other, important Higgs decay
channels at the Tevatron and at the LHC, $h \rightarrow b\bar{b}$ and
$h \rightarrow \tau^+ \tau^-$, are suppressed. First preliminary
studies in OPAL \cite{opal_newscans,eps03_bechtle} show that the $h
\rightarrow b\bar{b}$ suppressed region is kinematically out of 
reach at LEP2 while a significant part of the parameter space can be
excluded. A combination of the LEP data will address these new MSSM
benchmark scans in the near future, defining the reference point for
future MSSM Higgs boson searches at the LHC.

\subsection{CP-Violating MSSM Scenarios}
In the MSSM the Higgs potential is assumed to be invariant under CP
transformation at tree level. However, it is possible to break CP
symmetry in the Higgs sector by radiative corrections, especially by
contributions from third generation scalar quarks \cite{cpx}. Such a
scenario is theo\-retically attractive as it provides a possible
solution to the cosmic baryon asymmetry \cite{cpx_1}, while the
CP-violating effects predicted by the SM are too small to account for
it.

For the first time the LEP Higgs searches have also been interpreted
in the CP-violating MSSM scenario \cite{opal_newscans}. In such a
scenario the three neutral Higgs bosons, $H_i (i = 1, 2, 3)$ are
mixtures of the CP-even and CP-odd Higgs fields. Consequently, they
all couple to the Z boson and to each other, and these couplings may
be widely different from those of the CP-conserving scenario. In the
CP-violating scenario the Higgsstrahlung processes $e^+e^- \rightarrow
H_i H_j \; (i \ne j)$ may all occur, with widely varying cross
sections. In large domains of the model parameters, the lightest Higgs
boson $H_1$ may escape detection, although it has a predicted mass
that is well within the LEP range, since its coupling to the Z boson
is too weak for detection; on the other hand, the other two Higgs
boson masses can be out of reach or also may have small cross
sections. As a result the limits on MSSM parameters advertised so far
for the CP-conserving scenario can be invalidated.  The decay
properties of the Higgs bosons, while being quantitatively different
in the two scenarios, maintain a certain similarity. Since Higgs
bosons, in general, couple to mass, the largest branching ratios are
those to $b\bar{b}$ and $\tau \tau$ pairs. If kinematically allowed,
the decays $h \rightarrow AA$ (CP-conserving scenario) or $H_2
\rightarrow H_1 H_1$ (CP-violating scenario) would occur and could
even be dominant decays.

\begin{figure}[ht]
\centerline{
\resizebox{0.49\textwidth}{!}{%
  \includegraphics{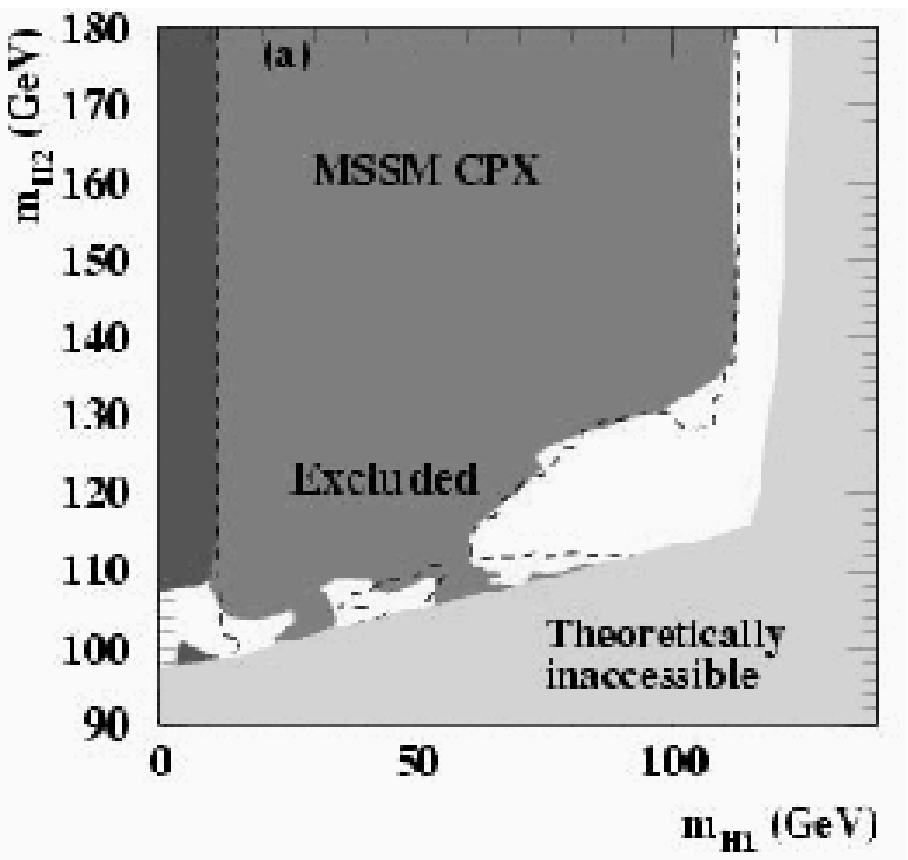} 
  \includegraphics{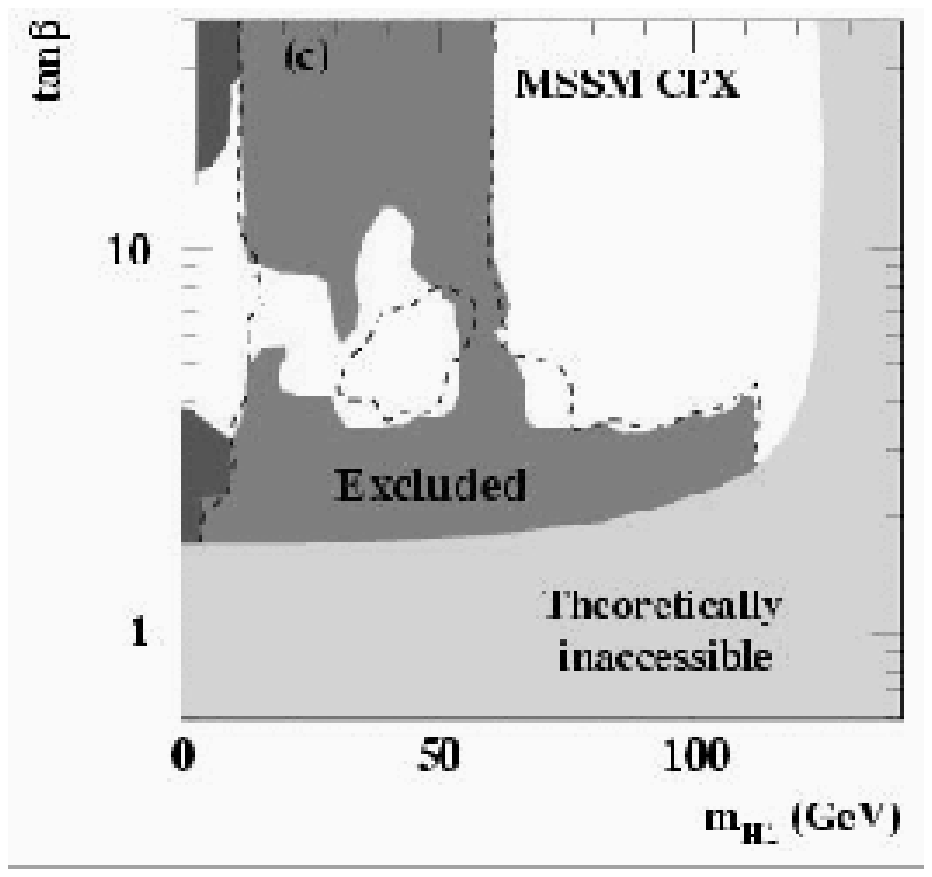} 
}
}
\caption{Preliminary OPAL results on exclusion areas in the 
         CP-violating MSSM. The observed exclusion region is medium
         shaded (green), the expected excluded region is indicated by
         the dashed line and the theoretically inaccessible region is
         light shaded (yellow). Regions excluded by Z width
         constraints or by decay mode independent searches are dark
         shaded (red).}
\label{fig:19}       
\end{figure}

Figure~\ref{fig:19} shows the preliminary exclusion results of the
Higgs search and interpretation in the CP-violating MSSM. In the area
of heavy $m_{H_2}$ the lighter $H_1$ resembles the SM Higgs boson with
very little effect from CP-violation (limit on $m_{H_1} > 112 \;\rm
GeV$). In the region below $m_{H_2} \approx 130 \;\rm GeV$ CP
violating effects play a major role. Exclusion is obtained for $\tan
\beta < 3.2$ and $m_{H_1} < 112 \;\rm GeV$ in the SM like regime. For 
$4 < \tan \beta < 10$ $ZH_2$ production is dominant. The large
difference between the expected and observed exclusion regions in the
area of $4 < \tan \beta <10$ is mainly due to a less than $2\,\sigma$
excess of events. For light $m_{H_1} < 50\;\rm GeV$ there are regions
expected to be unexcluded, due to the dominant $ZH_2 \rightarrow Z H_1
H_1$ production with relatively large $m_{H_1}$, yielding broadened
resolutions and therefore reduced sensitivity. 

CP-violating MSSM models will also be subject of the upcoming final
combination of LEP data by the LEP-Higgs working group.

\section{Search for Charged Higgs Bosons}
Extensions to the minimal Standard Model contain more than one Higgs
doublet. In particular, models with two complex Higgs doublets (2HDM)
predict two charged Higgs bosons $H^\pm$.  In 2HDM type-II the ratio
of the charged Higgs couplings to down- and up-type fermions is given
by the ratio of the vacuum expectation values for the two doublets. At
born level the charged Higgs mass has to be larger than the $W$-boson
mass. But radiative corrections can reduce the charged Higgs mass
below this threshold.

At LEP the search for the charged Higgs boson in $e^+e^- \rightarrow
H^+H^-$ is performed in the three decay channels $H^+H^- \rightarrow
\tau^+ \nu_\tau \tau^- \bar{\nu}_\tau$, $H^+H^- \rightarrow c \bar{s} \tau^-
 \bar{\nu}_\tau$ and $H^+H^- \rightarrow c\bar{s} \bar{c} s$, assumed
 to be the only possible decays.

\begin{figure}[ht]
\centerline{
\resizebox{0.49\textwidth}{!}{%
  \includegraphics{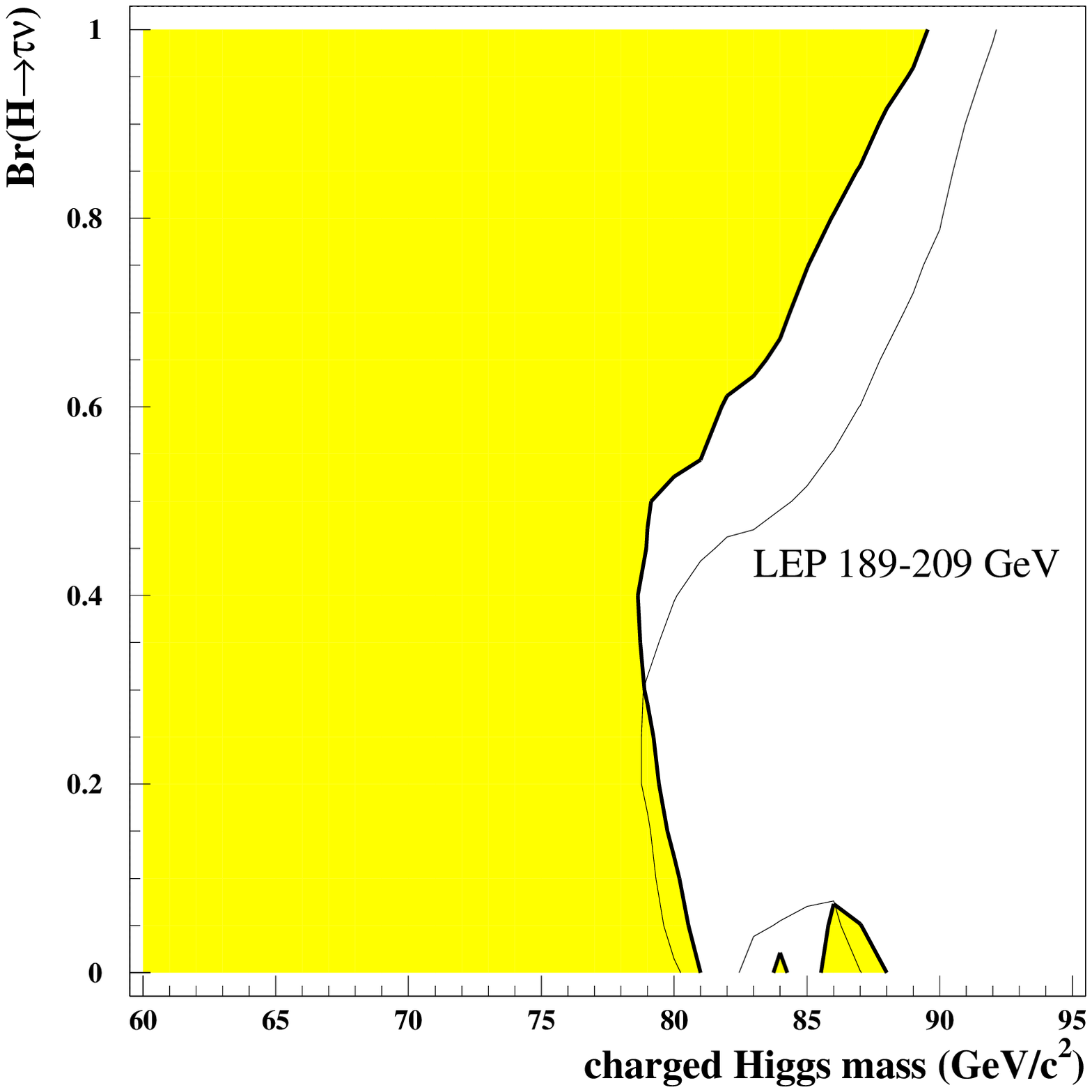}
  \includegraphics{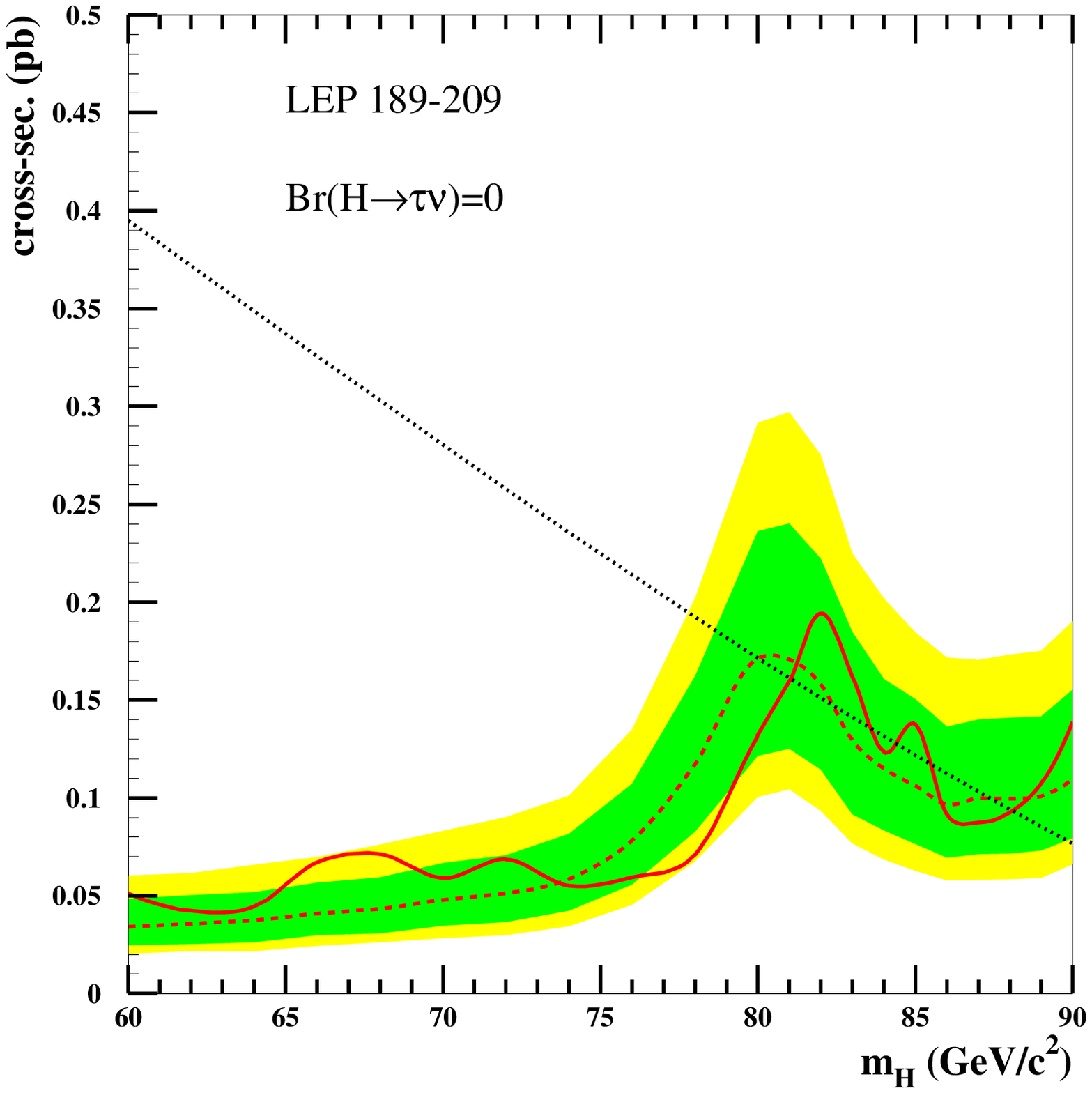}
}
}
\caption{Left: Preliminary LEP-combined exclusion region in the plane of 
         $Br(H^\pm \rightarrow \tau \nu)$ and $m_{H^\pm}$, using
         $\approx 2.5\;\rm fb^{-1}$. Right: Expected cross section
         for $e^+e^- \rightarrow H^+H^-$ production (dotted line),
         observed cross section limit (full line) and expected cross
         section limit along with 1 and $2\,\sigma$ error band as a
         function of the charged Higgs mass.}
\label{fig:20}       
\end{figure}

Figure~\ref{fig:20} shows the preliminary LEP-combined results
\cite{leph_chargedh} on the exclusion region in the plane of $Br(H^\pm
\rightarrow \tau \nu)$ and $m_{H^\pm}$ and , using $\approx 2.5\;\rm
fb^{-1}$. Also shown is the expected signal cross section along with
the expected and observed cross section limit as a function of the
charged Higgs mass for $Br(H^\pm \rightarrow \tau \nu) = 0$. The large
$W$-back\-ground at about $81\;\rm GeV$, which also defines the lower
mass bound can clearly be seen. At $m_{H^+} \approx 67\;\rm GeV$ a
$\sim 2\,\sigma$ excess is observed, which comes mainly from the L3
data. Within L3 this excess of events has been observed for data from
all centre-of-mass energies in all years, summing up to a $\sim
4.6\;\rm \sigma$ excess in the year 2000 \cite{leph_chargedh}.

\begin{figure}[ht]
\centerline{
\resizebox{0.24\textwidth}{0.20\textwidth}{%
  \includegraphics{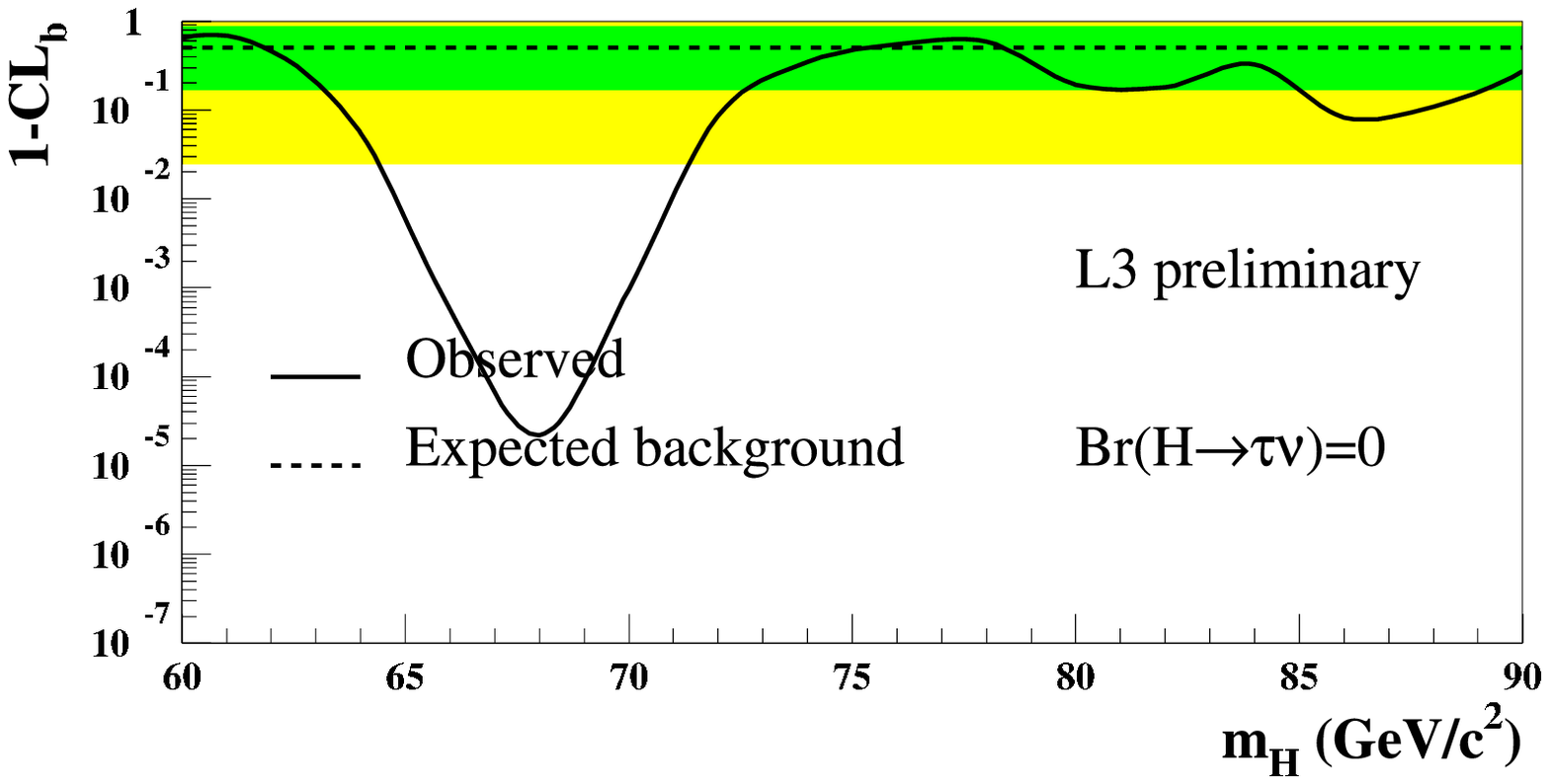}}
\resizebox{0.24\textwidth}{0.20\textwidth}{%
  \includegraphics{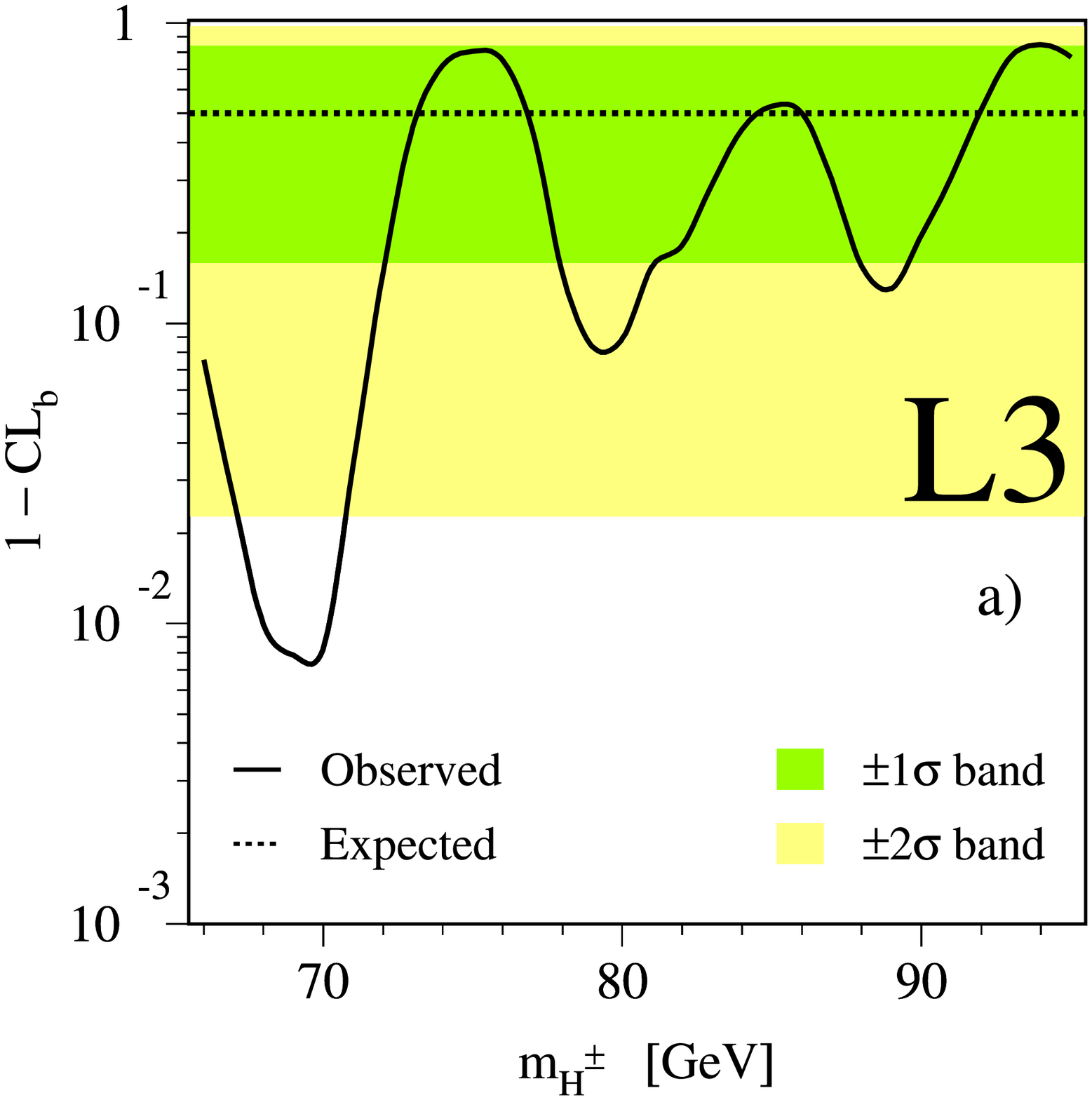}
}
}
\caption{Left: Confidence level $1 - CL_b$ for background-only hypothesis
         for charged Higgs L3 data from \cite{leph_chargedh}
         (year 2000). The observed values are shown as solid line, the
         expected value of 0.5 as dashed line and the $1$- and
         $2-\sigma$ band as shaded regions. An excess of about
         $4.6\,\sigma$ is observed at $m_{H^\pm} \approx 67\;\rm GeV$.
         Right: Confidence level $1 - CL_b$ for background-only
         hypothesis for latest charged Higgs L3 data from
         \cite{l3_chargedh_eps03}. The excess has been reduced to
         about $2.4\,\sigma$.}
\label{fig:21}       
\end{figure}

Figure~\ref{fig:21} shows the confidence level $1 - CL_b$ for the
background-only hypothesis as a function of the charged Higgs mass for
the L3 data. The large excess of events, resulting in a strong dip of
the confidence level at $m_{H^\pm} \approx 67\;\rm GeV$ can be clearly
seen.

In the meantime L3 has reanalysed their data using the latest detector
calibration and using the hadronic $W$-mass for cross checks
\cite{l3_chargedh_eps03}. In doing so the excess in the new preliminary 
results has been largely reduced to about $2.4\; \sigma$ (see
Figure~\ref{fig:21}), resolving the long-standing puzzle on this
low-mass excess. The final combination will be performed by the
LEP-Higgs working group in the near future.

OPAL has also interpreted a measurement of the ratio of $\tau$-decay
branching ratios to muons and electrons in the context of 2HDM type-II
and extracted mass limits \cite{opal_tau_chargedhiggs}. The ratio of
branching ratios $Br(\tau^- \rightarrow \mu^- \bar{\nu}_\mu \nu_\tau)
/ Br(\tau^- \rightarrow e^- \bar{\nu}_e \nu_\tau) = 0.9726 \left( 1 +
4 \eta {m_\mu}/{m_\tau} \right)$ allows a measurement of the
Michel parameter $\eta$. A non-zero value of $\eta$ may imply the
presence of scalar couplings. In the 2HDM type-II the relation $\eta =
-\, \frac{m_\tau m_\mu}{2} \left( {\tan \beta} / {m_{H\pm}}
\right)^2$ results in sensitivity of this measurement to $\tan \beta$ 
and the charged Higgs mass. The OPAL measurement yields a limit of
$M_{H^\pm} > 1.28\, \tan \beta\,\rm GeV$, complementing an earlier
limit of $M_{H^\pm} > 1.89\, \tan \beta\,\rm GeV$, obtained from
studies of $b \rightarrow \tau^- \bar{\nu}_\tau X$ decays.

\begin{figure}[ht]
\centerline{
\resizebox{0.49\textwidth}{!}{%
  \includegraphics{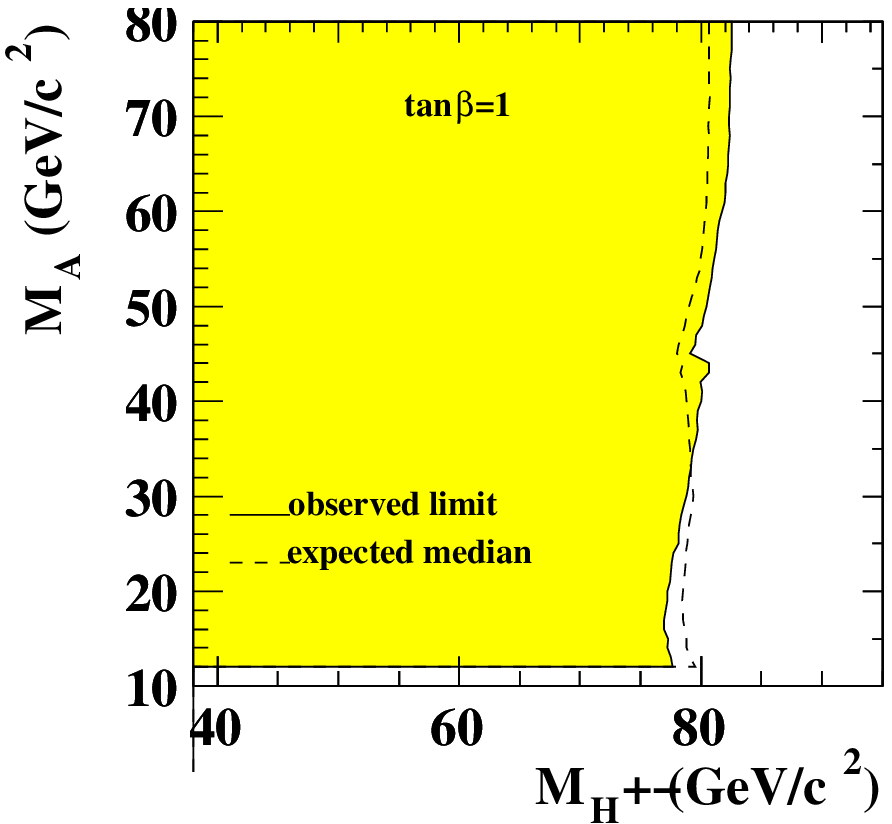}
  \includegraphics{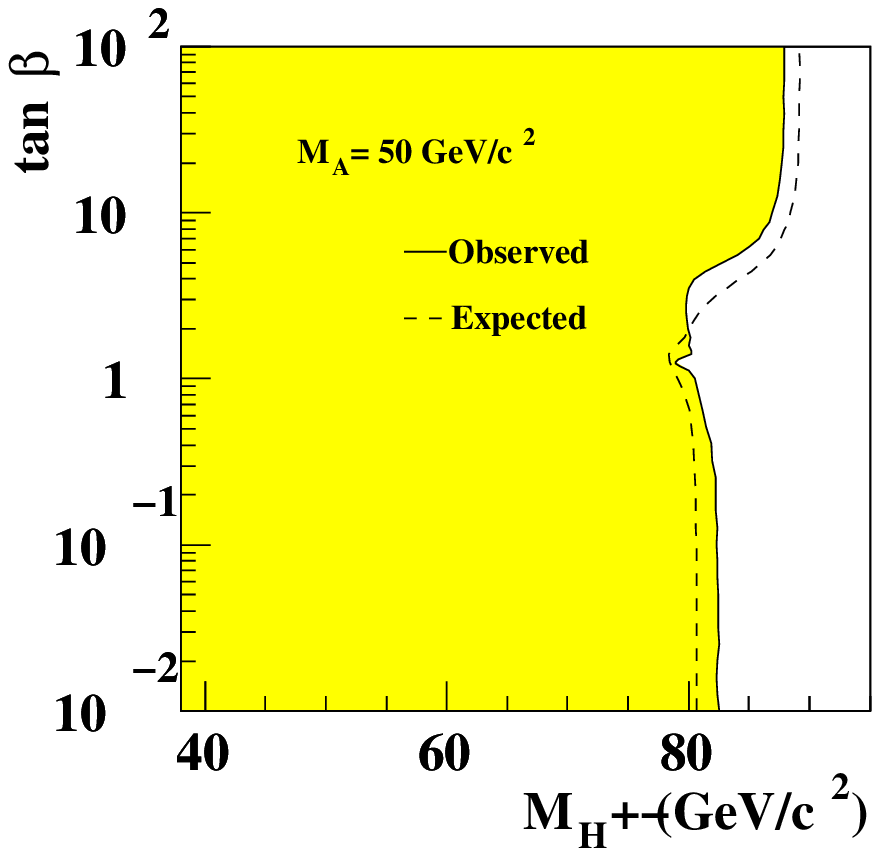} } }
\caption{Typical observed and expected exclusion regions from 
         preliminary DELPHI results in the 2HDM type-I.}
\label{fig:22}       
\end{figure}

DELPHI has also performed a search for charged Higgs bosons in the
context of 2HDM type-I \cite{delphi_2hdm_i}, where at $\tan \beta > 1$
the decay $H^\pm \rightarrow W^*A$ becomes allowed or even dominant,
depending on the choice of the CP-odd Higgs mass $m_A$. The decays of
the $W$ to leptons and hadrons are considered while the $A$
predominantly decays to $b\bar{b}$ in the parameter region
considered. These new decay channels are combined together with the
charged Higgs decays to $c \bar{s}$ and $\tau \nu$ and yield a mass
limit of $m_{H^\pm} > 76.6\;\rm GeV$, independently of $\tan \beta$
for $m_A > 12\;\rm GeV$ (Figure~\ref{fig:22}).

Further details and discussion on the status of searches for charged
Higgs bosons can be found in \cite{eps03_cuevas}.

\section{Fermiophobic Higgs Searches}
In 2HDM's without explicit CP-viola\-tion the couplings of the Higgs
doublets to fermions could be realized in different ways, one
possibility is that only one of the doublets couples to fermions. The
coupling of the lightest CP-even boson to a fermion pair is then
proportional to $\cos \alpha$. If $\alpha = \pi/2$ this coupling
vanishes and $h^0$ becomes a fermiophobic Higgs: it will decay to
pairs of other Higgs bosons or massive gauge bosons when kinematically
allowed, or to two photons in a large region of the parameter space.

The LEP collaborations have searched for fermiophobic Higgs bosons in
the $h \rightarrow \gamma \gamma$ and $h \rightarrow WW$ channels in
the Higgsstrahlung production $e^+e^- \rightarrow ZH$
\cite{lhwg_fermioph}. Lower mass bounds of $m_h > 109.7\;\rm GeV$ (observed) and
$m_h > 109.4\;\rm GeV$ (expected) have been obtained. Corresponding
searches at the Tevatron using Run-I or Run-II data presently have a
sensitivity of up to $\sim 80\;\rm GeV$ \cite{tevatron_fermioph}.

\begin{figure}[ht]
\centerline{
\resizebox{0.49\textwidth}{!}{%
  \includegraphics{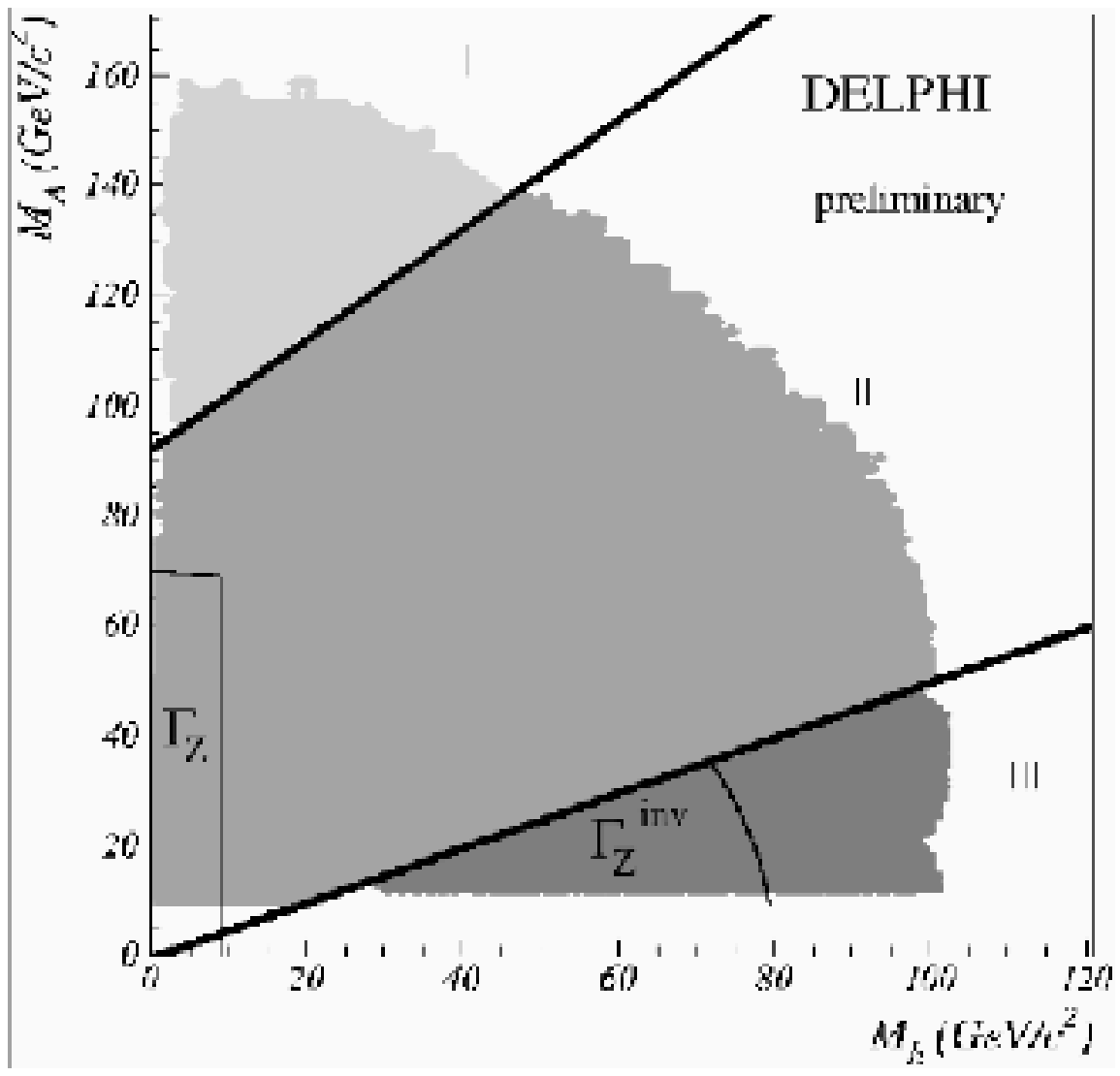}
  \includegraphics{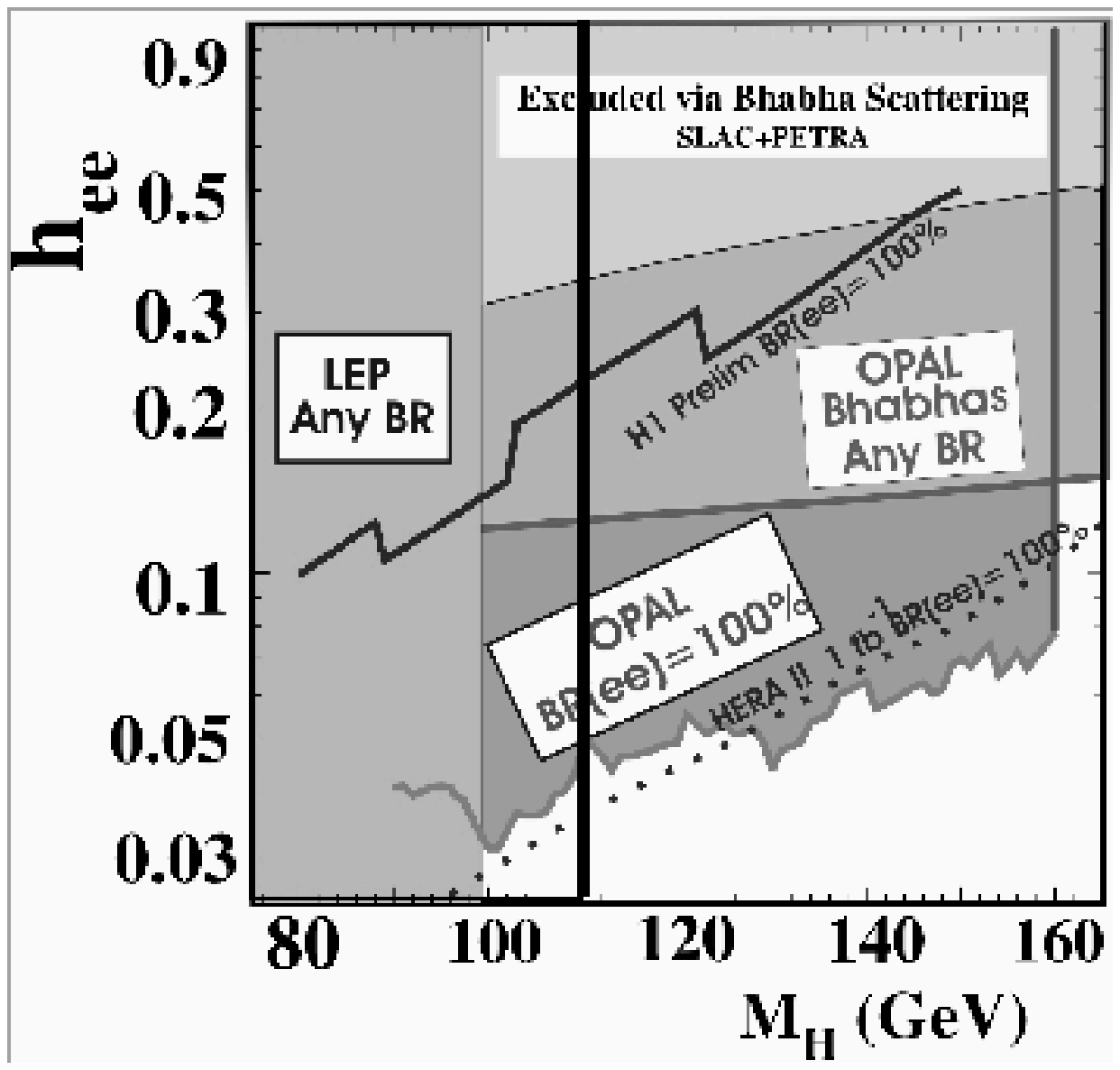}}}
\caption{Left: The shaded areas correspond to 
         the regions excluded for all values of $\delta$. The plot is
         devided in regions according to the dominant decay modes of
         $h$ and $A$; 
         right: Constraints on doubly-charged Higgses in the plane
         of Yukawa coupling and doubly-charged Higgs mass.}
\label{fig:23}       
\end{figure}

DELPHI has also searched for $e^+e^- \rightarrow hA$ with subsequent
decays to $h \rightarrow \gamma \gamma$ and $A \rightarrow b\bar{b}$
or $A\rightarrow hZ$ \cite{delphi_fermioph}. Also results for $h
\rightarrow AA$ and for long-lived A are considered. In general 2HDM,
the main mechanisms for the production of neutral Higgs bosons at LEP
are $e^+e^- \rightarrow hZ$ and $e^+e^- \rightarrow hA$, both
proceeding via $Z$ exchange. The two processes have complementary
cross sections proportional to $\sin^2 \delta$ and $\cos^2 \delta$,
respectively, where $\delta = \beta - \alpha$.

The combination of the results on $hZ$ and $hA$ production is shown in
Figure~\ref{fig:23}. The upper limits on $\sin^2 \delta$ for a given
$M_h$ and the upper limits on $\cos^2 \delta$ for a given $(M_h, M_A)$
pair are combined to exclude the $(M_h, M_A)$ pair for all $\delta$
values. $h$ masses up to $\approx 100\;\rm GeV$ can be excluded while
the sensitivity in $M_A$ reaches up to $\approx 160\;\rm GeV$.

Further discussionss on fermiophobic Higgs searches can be found in
\cite{eps03_fermioph}.

\section{Search for Doubly Charged Higgs Bosons}
Some theories beyond the Standard Model predict the existence of
doubly-charged Higgs bosons, $H^{\pm \pm}$ with $m_{H^{\pm
\pm}}$ ${\sim 100\;\rm GeV}$, including in particular Left-Right symmetric
models and Higgs triplet models. It has been particularly emphasized
that heavy right-handed neutrinos together with the see-saw mechanism
to obtain light neutrinos can lead to a doubly-charged Higgs boson
with a mass accessible to current and future colliders. Doubly-charged
Higgs bosons would decay into like-signed lepton or vector boson
pairs, or with a negligible branching fraction to a $W$-boson and a
singly-charged Higgs boson.  For masses less than twice the $W$-boson
mass, they would decay predominantly into like-signed leptons.

Limits on the Yukawa couplings exist from muonium conversion
experiments $(\mu^+ e^- \rightarrow \mu^- e^+)$ $\sqrt{h_{ee} h_{\mu
\mu}} \le 7.6 \cdot 10^{-3}\;\rm GeV^{-1}\, M_H$ and from 
avoiding large contributions to $(g-2)_\mu$, $h_{\mu \mu} \le 5.0 \cdot
10^{-3} \;\rm GeV^{-1}\, M_H$.

\begin{table}[ht]
\caption{Summary of the HERA searches for high mass multi-electron events.}
\label{tab:7}       
\centerline{
\begin{tabular}{cccc}
\hline\noalign{\smallskip}
experim. & selection & Data & SM expect. \\
\noalign{\smallskip}\hline\noalign{\smallskip}
H1   & 2 e with $M > 100\;\rm GeV$ & 3 & $0.30 \pm 0.05$ \\
H1   & 3 e with $M > 100\;\rm GeV$ & 3 & $0.23 \pm 0.04$ \\
ZEUS & 2 e with $M > 100\;\rm GeV$ & 2 & $0.77 \pm 0.08$ \\
ZEUS & 3 e with $M > 100\;\rm GeV$ & 0 & $0.34 \pm 0.09$ \\
\noalign{\smallskip}\hline
\end{tabular}}
\end{table}

In H1 a search for high mass multi-electron events has yielded an
excess of events at high masses \cite{hpp_h1}, as summarized in
Table~\ref{tab:7}. This observation was initially interpreted as a
possible indication for single production of a doubly-charged
Higgs. However, in a refined analysis, optimized for doubly-charged
Higgs production, one candidate event was selected with an expected SM
background of 0.34 events. Also no excess was observed by ZEUS in the
multi-electron channel and no excess was found by H1 or ZEUS in the
multi-muon channel. Therefore it is consi\-dered to be unlikely that
the H1 high-mass multi-electron events originate from doubly-charged
Higgs production.

Figure~\ref{fig:23} summarizes the constraints on doubly-
\linebreak charged
Higgs bosons in the plane of Yukawa coupling and doubly-charged Higgs
mass. The pair production searches at LEP managed to exclude
doubly-charged Higgs bosons up to the kinematic limits of $\sim
\sqrt{s}/2$ \cite{hpp_pair}. The single production limits of 
doubly-charged Higgses from H1 is shown to depend on the Yukawa
coupling and is at masses around $100 -140\;\rm GeV$ slightly stronger
than the Bhabha scattering limits from SLAC and PETRA
\cite{hpp_slac}. Between $100 - 120 \;\rm GeV$ the effect of the
candiate events can be seen. OPAL's analysis on single production of
doubly-charged Higgs \cite{hpp_single_opal} sets strong limits on the
Yukawa coupling for masses between 100 and $\sim 160\;\rm GeV$ which
are already equivalent to the expected sensitivity from $1\;\rm
fb^{-1}$ of HERA-II running. Analysis on Bhabha scattering by OPAL and
L3 \cite{hp_opal_bhabha} provides relatively strong limits on the
Yukawa coupling up to very high masses.

With analyses based on {\bf $91-107\;\rm pb^{-1}$} CDF and D{\O} are
already now able to exclude doubly-charged Higgs bosons with masses up
to $\sim 116\;\rm GeV$ \cite{hpp_tev_ii}. With the expected luminosity
of $\sim 2\;\rm fb^{-1}$ per experiment in Run-IIA the Tevatron
exclusion limit, which is independent of the Yukawa coupling due to
the pair production mechanism, is expected to increase up to $150 -
200\;\rm GeV$, covering the entire plane shown in Figure~\ref{fig:23}.

\section{Conclusion}
The Standard Model is in various aspects incomplete, giving us strong
indications that physics beyond the Standard Model is to be
expected. However, we are presently groping in the dark on where and
how we will find it. Fundamental questions on the origin and role of
the electroweak, the TeV and the Planck scale remain
unanswered. Similarly the origin of the observed particle structure in
the lepton and quark sector persists to be a puzzle.

We have a wealth of data in hand, allowing for model-independent
searches, searches within numerous models, measurements of rare
decays, precision electroweak measurements, and we expect increasing
sensitivity from the Tevatron Run-II, HERA-II and eventually from the
LHC. So far no significant deviation from the Standard Model
expectation has been observed.  Nevertheless the strongest indication
for new or so far unobserved physics at or from the TeV scale comes
from the Higgs sector and electroweak symmetry breaking, where a light
Higgs boson appears to be favoured by the presently available dataset.
The search goes on at HERA, the Tevatron, and in the future at the LHC
or even a linear collider.

\section*{Acknowledgements}
I would like to thank the organisers of the HEP2003 Europhysics
Conference in Aachen, Germany, for their hospitality. I am grateful to
many colleagues for their help in preparing this review and for
carefully reading this manuscript, in particular to F.~Bedeschi,
C.~Berger, V.~B\"u\-scher, H.~Drei\-ner, A.~Frey, E.~Gallo,
G.~Landsberg, P.~Newman, C.~Rembser, A.~Sch\"oning, T.~Wengler, and
P.~Igo-Kemenes. I do also thank DESY, CERN and Fermilab for the help
and support I received as a member of the ZEUS, OPAL, ATLAS and D{\O}
Collaborations.

%

\begin{thebibliography}{}
%
%
\bibitem{eps_pippa} EPS'03, P.~Wells, Experimental Tests of the Standard Model.
\bibitem{lhc_lc_summaries} EPS'03, S.~Arcelli, Search for $H/A \rightarrow \mu \mu$ at the LHC; 
                           EPS'03, E.~Ros, Prospects for Little Higgs Models at the LHC;
                           EPS'03, G.~Weiglein, Interplay between the LHC and a Linear Collider 
                                                in Searches for New Physics; 
                           EPS'03, S.~Hesselbach, New ideas on SUSY Searches at Future Linear Colliders.
\bibitem{exferm_models} K.~Hagiwara, S.~Komamiya and D.~Zeppenfeld, 
                        Z.Phys. C29 (1985) 115;
                        U.~Bauer, M.Spira and P.M.~Zerwas, 
                        Phys.Rev. D42 (1990) 815;
                        F.~Boudjema, A.~Djouadi and J.L.~Kneur, 
                        Z.Phys. C57 (1993) 425.

\bibitem{estar_lep_hera} LEP working group on exotics, LEP Exotica WG 2001-02;
                         ZEUS Collaboration,
                         DESY-01-132, Phys. Lett. B 549 (2002) 32;
                         H1 Collaboration, DESY-02-096, 
                         Phys. Lett. B548 (2002) 35;
                         H1 Collaboration, DESY
                         01-145, Phys. Lett. B 525 (2002) 9.
\bibitem{estar_tev_lhc}  E.~Boos et al., hep-ph/011034; 
                         O.J.P.~Eboli et al., hep-ph/0111001.

\bibitem{qstar_search_exp} CDF Collaboration,
                         Phys.Rev.Lett. 72 (1994) 3004; DELPHI
                         Collaboration, CERN-EP/98-169;
                         ZEUS Collaboration; 
                         Phys. Lett. B 549 (2002) 32;
                         H1 Collab., Eur. Phys. J. C17 (2000) 567.

\bibitem{eps2003_excited_fermions}
                         EPS'03, E.~Sanchez, Search for Excited Fermions.
\bibitem{brw_model}      W.~Buchm\"uller, R.~R\"uckl and D.~Wyler, 
                         PHys.Lett.B 191, 442 (1987); erratum in Phys.Lett.B 448, 320
                         (1999).
\bibitem{eps03_leptoquark} EPS'03, A.~Zarnecki, Search for Leptoquark Production and 
                          Lepton Flavour Violation and references therein.

\bibitem{hera_highpi_met} H1 Collab., DESY-03-132, Submitted to Eur.Phys.J;
                          ZEUS Collaboration; DESY-03-012,
                          Phys.Lett.B 559 (2003) 153.


\bibitem{kuze_sirois}     M.~Kuze and Y.~Sirois, Search for Particles and 
                          Forces Beyond the Standard Model at HERA $ep$ and 
                          Tevatron $p\bar{p}$ Colliders, hep-ex/0211048.




\bibitem{lep_t}          ALEPH Collaboration, Phys.Lett.B 543 (2002) 173;
                         DELPHI Collaboration, DELPHI 2003-042 CONF 662;
                         L3 Collaboration, Phys.Lett.B 549 (2002) 290;
                         OPAL Collaboration, Phys.Lett.B 521 (2001) 181.

\bibitem{cdf_t}          CDF Collaboration, Phys.Rev.Lett. 80, 2525 (1998).
\bibitem{fcnc_tevii}     T.~Han et al., hep-ph/9603247, 
                         T.~Han et al. hep-ph/9506461.
\bibitem{fcnc_heraii}    Estimate of the HERA-II sensitivity to 
                         $\kappa_{tu\gamma} - v_{tuZ}$ in the absence
                         of a signal; D.~Dannheim, private
                         communication.
\bibitem{eps03_singletop} EPS'03, J.~Ferrando, Single Top Production via FCNC.
\bibitem{add}            N.~Arkani-Hamed, S.~Dimopoulos and G.~Dvali, 
                         Phys.Lett. B 429, 263 (1009); 
                         Phys.Lett.B 436, 257 (1998).
\bibitem{rs}             L.~Randall and R.~Sundrum, Phys.Rev.Lett. 83, 3370 
                         (1999); Phys.Rev.Lett. 83, 4690 (1999).
\bibitem{grw}            G.F.~Giudice, R.~Rattazzi and J.D.~Wells, 
                         Nucl.Phys. B 544, 3 (1999).
\bibitem{shortrange_gravity} J.~Long, J.~Price, hep-ph/0303057 and 
                         references therein.  

\bibitem{opal_radion}    The OPAL Collaboration, PN 526.
\bibitem{l3_branons}     The L3 Collaboration, L3 Internal Note 2814.

\bibitem{more_eps03_ed} EPS'03, I.~Antoniadis, Physics with large extra dimensions;
                        EPS'03, S.~Mele, Current experimental bounds on extra dimensions;
                        EPS'03, L.~Vacavant, Search for extra dimensions with ATLAS at LHC;
                        EPS'03, F.~del Aguila, Extra Dimensions with Brane Localized Kinetic Terms;
                        EPS'03, F.~Feruglio, Extra Dimensions in Particle Physics;
                        EPS'03, M.~Sanders, Search for Extra Dimensions at Hadron Colliders.
\bibitem{eps03_susy}    EPS'03, P.~Azurri, Gaugino Searches and SUSY Models Constraints;
                        EPS'03, M.~Wegner, Searches for MSSM/MSUGRA at the Tevatron-II;
                        EPS'03, I.~Trigger, Searches for sfermions;
                        EPS'03, S.~Hesselbach, New ideas on SUSY Searches at Future Linear Colliders;
                        EPS'03, T.~Nunnemann, Search for SUSY in GMSB and AMSB Models;
                        EPS'03, C.~Schwanenberger,  Search for RPV SUSY.

\bibitem{eps03_rott}    EPS'03, C.~Rott, Search for top and bottom squarks.
\bibitem{lep_neutralino_limit} LEP SUSY working group, LEPSUSYWG/01-07.1.
\bibitem{lsp_cdm}       D.~Hooper, T.~Plehn, hep-ph/0212226, 
                        Bottino et al., Phys.Rev.D 67, 063519 (2003).
\bibitem{lsp_sn1987}    H.~Dreiner et al., hep-ph/0304289.
\bibitem{p_higgs}       P.W.~Higgs, Phys.Lett 12 (1964) 132;
                        {\it idem}, Phys.Rev.Lett. 13 (1964) 508;
                        {\it idem}, Phys.Rev. 145 (1966) 1156;
                        F.~Englert and R.~Brout, Phys.Rev.Lett. 13 (1964) 321;
                        G.S.~Guralnik, C.R.~Hagen and T.W.B.~Kibble, 
                        Phys.Rev.Lett. 13 (1964) 585.
\bibitem{lepew_hfit}    The LEP Collaborations ALEPH, DELPHI, L3 and OPAL, 
                        the LEP Electroweak Working Group and the SLD
                        Heavy Flavour Group, {\it A Combination of
                        Preliminary Electroweak Measurements and
                        Constraints on the Standard Model}, CERN-EP/
                        2003-02, hep-ex/0312023.
\bibitem{leph_smh_final} The LEP Collaborations ALEPH, DELPHI, L3 and OPAL, 
                        the LEP Working Group for Higgs Boson Searches,
                        {\it Search for the Standard Model Higgs Boson at LEP},
                        CERN-EP/2003-011, Phys. Letts. B565 (2003) 61-75.
\bibitem{higgs_sensitivity_ii} B.~Klima et al., Tevatron Higgs Sensitivity Working Group,
                        Results of the Tevatron Higgs Sensitivity Study, FERMILAB-PUB-03/320-E.
\bibitem{susyhiggs_runii_study} M.~Carena et al., Report of the Tevatron 
                        Run 2 SUSY/Higgs Working Group,
                        hep-ph/0010338.  
\bibitem{runi_top_mass} J.~Estrada, Optimal use of Information for Measuring $M_t$
                        in Lepton+jets $t\bar{t}$ Events,
                        hep-ph/0302031.
\bibitem{eps03_varelas} EPS'03, N.~Varelas, Electroweak and Higgs Physics with D{\O}.
\bibitem{lhc_sm_higgs}  ATLAS Collaboration, Detector and Physics Performance Technical 
                        Design Report, CERN/LHCC/99-14 (1999); 
                        CMS Collaboration, CMS Technical proposal, CERN/LHCC 94-38, CERN (1994).
\bibitem{wbf_higgs_zepp1} D.L.~Rainwater and D.~Zeppenfeld, J.~High Energy Phys. 
                          12 (1997) 4, hep-ph/9712271;
                          D.L.~Rainwater and D.~Zeppenfeld, Phys. Rev. 
                          D 60 (1999) 113004, hep-ph/9906218.
\bibitem{wbf_higgs_zepp2} D.L.~Rainwater, D.~Zeppenfeld and K.~Hagiwara, Phys. Rev. 
                          D 59 (1999) 14037, hep-ph/9808468;
                          T.~Plehn, D.L.~Rainwater and D.~Zeppenfeld, Phys. Rev. 
                          D 61 (2000) 093005.
\bibitem{atlas_wbf_higgs} ATLAS Collaboration, Prospects for the Search for a Standard 
                          Model Higgs Boson at ATLAS using Vector Boson Fusion;
                          proceedings of Les Houches workshop 2002.

\bibitem{cms_wbf_higgs}  K.~Mazumdar for the CMS Collaboration, hep-ex/0308070.

\bibitem{eps03_duersen}  EPS'03, M.~Duersen, Standard Model Higgs Searches at CERN.
\bibitem{nlo_tth}        W.~Beenakker et al., Phys. Rev. Lett. 87, (2001) 201805
                         (hep-ph/0107081); Nucl. Phys. B 653, (2003) 151 (hep-ph/0211352);
                         L.~Reina, S.~Dawson, Phys. Rev. Lett. 87 (2001) 201804 
                         (hep-ph/0107101); L.~Reina, S.~Dawson, D.~Wackeroth, Phys. Rev. 
                         D 65 (2002) 053017 (hep-ph/0109066); S.~Dawson et al., Phys. 
                         Rev. D 67 (2003) 071503 (hep-ph/0211438); S.~Dawson et al., 
                         hep-ph/0305087.
\bibitem{nnlo_gg_fusion} R.~Harlander, W.B.~Kilgore, Phys.Rev.Lett. 88 (2002) 201801,
                         Phys.Rev. D68 (2003) 013001;
                         C.~Anastasiou, K.~Melnikov, Nucl.Phys. B646 (2002) 220;
                         V.~Ravindran, J.~Smith, W.~van Neerven, Nucl.Phys. B634 (2002) 247,
                         Nucl.Phys. B665 (2003) 325.
\bibitem{nlo_gg_gaga}    T.~Binoth et al., Phys. Rev. D 63 (2001) 114016 (hep-ph/0012191).
\bibitem{nlo_bh}         C.~Campbell et al., Phys. Rev. D 67 (2003) 095002 (hep-ph/0204093).
\bibitem{wh_zh}          M.L.~Ciccolini, S.~Dittmaier, M.~Kr\"amer, 
                         hep-ph/0306234. 
\bibitem{leph_mssm_cern} The LEP-Higgs Working Group, Searches for the Neutral Higgs Boson
                         of the MSSM: Preliminary Combined Results Using LEP Data Collected
                         at Energies up to 209 GeV, LHWG-Note 2001-04.
\bibitem{lhc_scans}      M.~Carena et al., Suggestions for Benchmark Scenarios for MSSM 
                         Higgs Boson Searches at Hadron Colliders, hep-ph/0202167.
\bibitem{opal_newscans}  OPAL Collaboration, PN524.
\bibitem{eps03_bechtle}  EPS'03, P.~Bechtle, Search for SUSY Neutral Higgs and interpretations.


\bibitem{cpx}            A.~Pilaftsis and C.E.~Wagner, Nucl. Phys. B 553 (1999) 3.
\bibitem{cpx_1}          M.~Carena et al., Nucl. Phys. B 599 (2001) 158.
\bibitem{leph_chargedh}  LEP-Higgs working group, Search for Charged Higgs bosons: 
                         Preliminary Combined Results Using LEP Data 
                         Collected at Energies up to 209 GeV, LHWG Note/2001-05.

\bibitem{l3_chargedh_eps03}  L3 Collaboration, L3 note 2817.
\bibitem{opal_tau_chargedhiggs} OPAL Collaboration, 
                         CERN-EP/2002-085; Phys. Lett. B 551 (2003) 35.
\bibitem{delphi_2hdm_i}  DELPHI Collaboration, DELPHI 2003-038 CONF 658.
\bibitem{eps03_cuevas}   EPS'03, J.~Cuevas, Search for (Singly and Doubly) charged Higgses.
\bibitem{lhwg_fermioph}  LEP-Higgs working group, Searches for Higgs Bosons Decaying into Photons: 
                         Combined Results from the LEP Experiments 
                         LHWG Note/2002-02.

\bibitem{tevatron_fermioph} CDF Collaboration, Phys. Rev. D 59 (1999) 092002;
                         D{\O} Collaboration, Phys. Rev. Lett. 82 (1999) 2244.
\bibitem{delphi_fermioph} DELPHI Collaboration, 
                         DELPHI 2003-043 CONF 663.
\bibitem{eps03_fermioph} EPS'03, D.~Baden, Fermiophobic Higgs Searches.
\bibitem{hpp_h1}         H1 Collab., A.~Aktas et al., Accepted by Eur.Phys.J.

\bibitem{hpp_pair}       DELPHI Collaboration, Phys. Lett. B 552 (2003) 127;
                         OPAL Collaboration, Phys. Lett. B 526 (2002) 221;
                         L3 Collaboration, L3 note 2818.

\bibitem{hpp_slac}       M.L.~Swartz, Phys. Rev. D 40 (1989) 1521.
\bibitem{hpp_single_opal} The OPAL Collaboration, 
                          CERN-EP-2003-041, Accepted by Phys.Lett.B.
\bibitem{hp_opal_bhabha} OPAL Collaboration, CERN-EP-2003-041, Accepted by Phys Lett. B.;
                         L3 Collaboration, CERN-EP/2003-060.
\bibitem{hpp_tev_ii}     CDF Collaboration, CDF note 6342,
                         D{\O} Collaboration, D{\O}-Note 4217.
                         


\end{thebibliography}
%
 
\end{document}